\documentclass[manuscript,screen]{acmart}

\AtBeginDocument{%
  \providecommand\BibTeX{{%
    \normalfont B\kern-0.5em{\scshape i\kern-0.25em b}\kern-0.8em\TeX}}}

\usepackage{algorithmic}
\usepackage{graphicx}
\usepackage{textcomp}
\usepackage{xcolor}
\usepackage{multirow,color, hyperref, url, graphics}
\usepackage{listings}
\usepackage{diagbox}
\usepackage{subfig}
\usepackage{balance}
\usepackage{adjustbox}
\usepackage{colortbl}
\usepackage{framed}
\usepackage{booktabs}
\usepackage{indentfirst}
\usepackage{makecell,tcolorbox}

\begin{document}

\title{Characterizing and Detecting WebAssembly Runtime Bugs}

\author{Yixuan Zhang}
\affiliation{%
  \institution{Peking University}
  \city{Beijing}
  \country{China}}
\email{zhangyixuan.6290@pku.edu.cn}

\author{Shangtong Cao}
\affiliation{%
  \institution{Beijing University of Posts and Telecommunications}
  \city{Beijing}
  \country{China}}
\email{shangtongcao@bupt.edu.cn}

\author{Haoyu Wang}
\affiliation{%
  \institution{Huazhong University of Science and Technology}
  \city{Wuhan}
  \country{China}}
\email{haoyuwang@hust.edu.cn}

\author{Zhenpeng Chen}
\affiliation{%
  \institution{University College London}
  \city{London}
  \country{UK}}
\email{zp.chen@ucl.ac.uk}

\author{Xiapu Luo}
\affiliation{%
  \institution{The Hong Kong Polytechnic University}
  \city{Hong Kong}
  \country{China}}
\email{csxluo@comp.polyu.edu.hk}

\author{Dongliang Mu}
\affiliation{%
  \institution{Huazhong University of Science and Technology}
  \city{Wuhan}
  \country{China}}
\email{dzm91@hust.edu.cn}

\author{Yun Ma}
\affiliation{%
  \institution{Peking University}
  \city{Beijing}
  \country{China}}
\email{mayun@pku.edu.cn}

\author{Gang Huang}
\affiliation{%
  \institution{Peking University}
  \city{Beijing}
  \country{China}}
\email{hg@pku.edu.cn}

\author{Xuanzhe Liu}
\affiliation{%
  \institution{Peking University}
  \city{Beijing}
  \country{China}}
\email{liuxuanzhe@pku.edu.cn}

\renewcommand{\shortauthors}{Yixuan Zhang, Shangtong Cao and Haoyu Wang, et al.}

\begin{abstract}
WebAssembly (abbreviated WASM) has emerged as a promising language of the Web and also been used for a wide spectrum of software applications such as mobile applications and desktop applications. These applications, named as WASM applications, commonly run in WASM runtimes. Bugs in WASM runtimes are frequently reported by developers and cause the crash of WASM applications. 
However, these bugs have not been well studied.
To fill in the knowledge gap, we present a systematic study to characterize and detect bugs in WASM runtimes. We first harvest a dataset of 311 real-world bugs from hundreds of related posts on GitHub. Based on the collected high-quality bug reports, we distill 31 bug categories of WASM runtimes and summarize their common fix strategies. Furthermore, we develop a pattern-based bug detection framework to automatically detect bugs in WASM runtimes. We apply the detection framework to five popular WASM runtimes and successfully uncover 53  bugs that have never been reported previously, among which 14 have been confirmed and 6 have been fixed by runtime developers.
\end{abstract}



\maketitle
\newcommand{\zhangyx}[1]{\textbf{\color{blue}{ZYX: #1}}}
\newcommand{\haoyu}[1]{{\color{red} HW: #1}}
\newcommand{\czp}[1]{{\color{red} ZP: #1}}
\newcommand{\mdl}[1]{{\color{cyan} DM: #1}}
\newcommand{\may}[1]{{\color{yellow} MY: #1}}
\newcommand{\xzl}[1]{\textbf{\color{violet}{(Xuanzhe: #1)}}}

\newcommand{\ie}{\mbox{\emph{i. e.,\ }}}

\section{Introduction}\label{sec:introduction}
WebAssembly (abbreviated WASM) has quickly emerged as a promising language of the Web in recent years~\cite{bringing}. WASM is a binary instruction specification \cite{bringing, slicing_of_wasm_binaries, newagain} for a stack-based virtual machine and provides developers with an equivalent textual format \cite{watfile} for reading, testing, learning instructions, and debugging, etc. Although WASM was initially proposed for Web applications \cite{wasim, arewethere}, it is moving fast towards a much wider spectrum of domains, including desktop applications \cite{wasm_non_web, WAVM}, mobile applications \cite{wasm_non_web}, IoT \cite{wasmiot, wasmachine2020}, blockchain \cite{wasm_blockchain, eos-vm, hera},  serverless computing \cite{wasm_non_web, wasm_serverless_edge}, and edge computing \cite{WASM_edge_computing, sledge}, etc..
To develop these applications (named WASM applications), developers can compile high-level programming languages to WASM binaries or convert the equivalent manually-written textual format to WASM binaries.
WASM binaries are commonly executed in WASM runtimes. A WASM runtime provides an efficient, memory-safe, sandboxed execution environment for WASM applications \cite{wasmdoc}. 
However, a great variety of WASM runtime specific bugs have been reported by developers, inevitably impeding the development of the WASM application ecosystem. 
Despite this, WASM runtime bugs have not been systematically studied by our community. Therefore, there is a general lack of an understanding of these bugs, including their root causes, fix patterns, and how to detect these bugs in emerging WASM runtimes.

\textbf{This Work.}
To fill in the knowledge gap, we present the first comprehensive study on characterizing and detecting bugs in WASM runtimes. We focus our study on three most popular and representative 
WASM runtimes, including wasmtime \cite{wasmtime}, wasmer \cite{wasmer}, and wasm-micro-runtime (WAMR) \cite{WAMR}. 
We first collect 903 bug-related posts from GitHub, a commonly-used data sources for studying software bugs, and make an effort to identify 311 real-world bugs of these WASM runtimes (see \textbf{\ref{sec:methodology}}).
Based on the collected bugs, we manually construct a taxonomy of 31 bug categories (see \textbf{\ref{sec:RQ1}}), indicating the diversity of WASM runtime bugs. Moreover, we summarize common fix patterns for each bug category (see \textbf{\ref{sec:RQ2}}). These empirical results provide a high-level categorization that can serve as a guide for developers to resolve common faults and for researchers to develop tools for detecting and fixing common WASM runtime bugs.

Furthermore, we develop a pattern-based bug detection framework based on the knowledge summarized from the bug taxonomy, to test the presence of bugs in WASM runtimes (see \textbf{\ref{sec:testsuite}}).
To evaluate the generalizability of our study, beyond the three analyzed WASM runtimes, we further consider two emerging WASM runtimes (\texttt{wasm3} and \texttt{WASMEdge}) for bug detection. We have successfully identified 53 previously-unknown bugs. We report these bugs to the developers of corresponding WASM runtimes. By the time of this writing, 14 bugs have been confirmed by the developers, and 6 of them have been fixed based on our suggestions. 

To summarize, this paper makes the following contributions:
\begin{itemize}
\item We conduct the first systematic study of bugs in WASM runtimes. We summarize common bug categories and their corresponding fix strategies. Our results can help understand and characterize bugs in WASM runtimes while shedding lights on future WASM related studies.

\item We develop a pattern-based bug detection framework based on the knowledge summarized from bug categories we created to automatically detect bugs in WASM runtimes. By applying the detection framework to real-world WASM runtimes, it shows that our proposed framework can effectively detect bugs and provide useful information to facilitate bug diagnosis and fixing.

\item We will make the scripts, datasets, and bug detector available to the research community for other researchers to replicate and build upon.

\end{itemize}

\section{Background}\label{sec:background}
\subsection{WASM binaries}
WASM is a low-level assembly-like language that is designed for efficient execution and compact representation. The WASM binary file is compact like Java class files and is saved with the \texttt{.wasm} suffix \cite{wasmbook}. The WASM specification defines a conceptual stack virtual machine for most WASM instructions to work on, performing numbers' pop and push and leaving the result on the stack. A pretty-printed textual format (i.e., \texttt{.wat}) \cite{watfile} is also provided for developers, which can be used to learn the syntax, understand the WASM module, test WASM program, optimize applications, debug code, and write WASM programs by hand, etc. 

Developers and users can use the wabt \cite{wabt} tool to translate WASM binaries to WASM textual format or vice versa.

\begin{figure}[htb]
    \centerline{\includegraphics[width=0.75\textwidth]{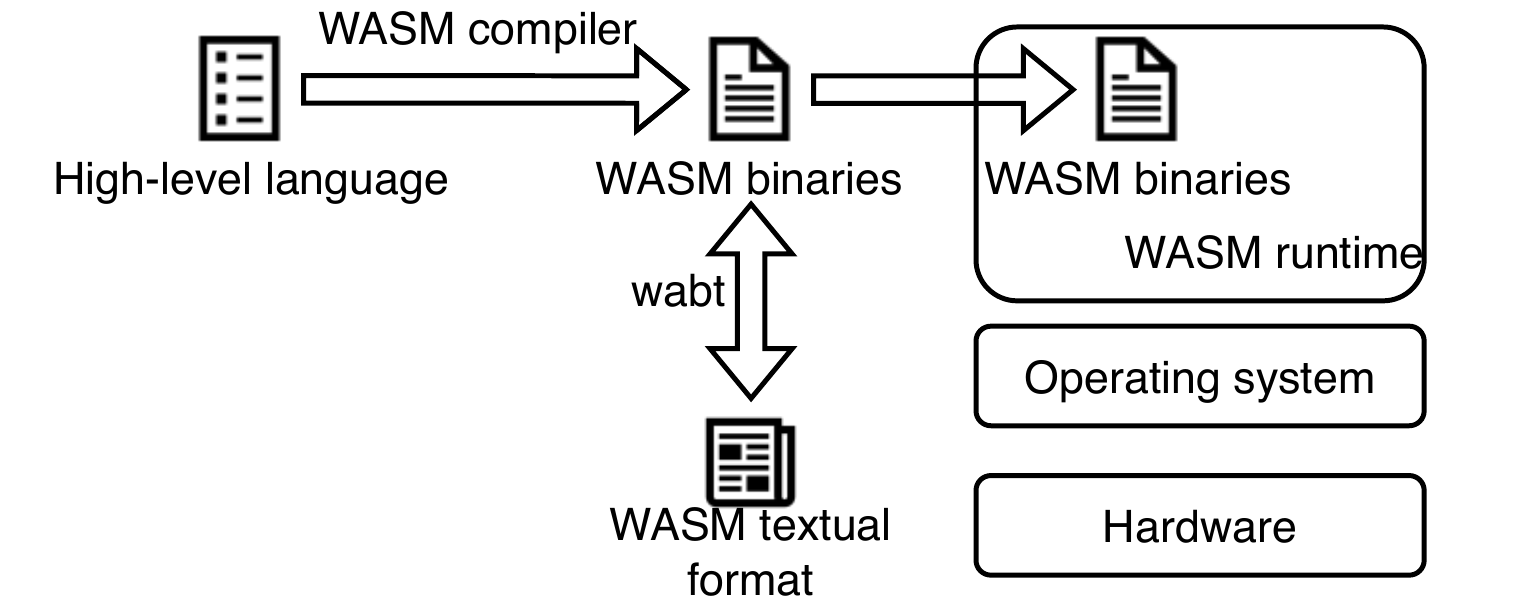}}
    \caption{The execution process of WASM binaries. 
    }
    \label{figure:workflow}
\end{figure}

\subsection{Execution of WASM binaries}
As a binary instruction format, WASM is designed as a portable compilation target for high-level programming languages \cite{wasmdoc}. As shown in Figure~\ref{figure:workflow}, developers can use WASM compilers to translate high-level language programs to WASM binaries. There are dozens of compilers available to compile different source language programs to WASM binaries, such as AssemblyScript, Emscripten, Rustc/WASM-Bindgen, etc \cite{empiricalstudy_wasmcompiler}.
WASM can be executed at native speed \cite{bringing} on a wide range of platforms. The tool for this critical process is a WASM runtime, an intermediate layer between the WASM binaries and the hardware platforms. A WASM runtime should consider the structure, operating system, and other differences between various platforms and provide a relatively secure execution environment for the WASM binaries. 
As shown in Figure ~\ref{figure:workflow}, developers can create applications in high-level languages, compile them into WASM binaries \cite{empiricalstudy_wasmcompiler, craneliftdoc}, and execute WASM binaries in WASM runtimes. Alternatively, they could develop simple WASM programs in the textual format, convert them to WASM binaries through \texttt{wabt} \cite{wabt}, and execute the binaries in WASM runtimes.

\subsection{WASM Runtime Architecture}

\begin{figure}[htb]
    \centerline{\includegraphics[width=0.75\textwidth]{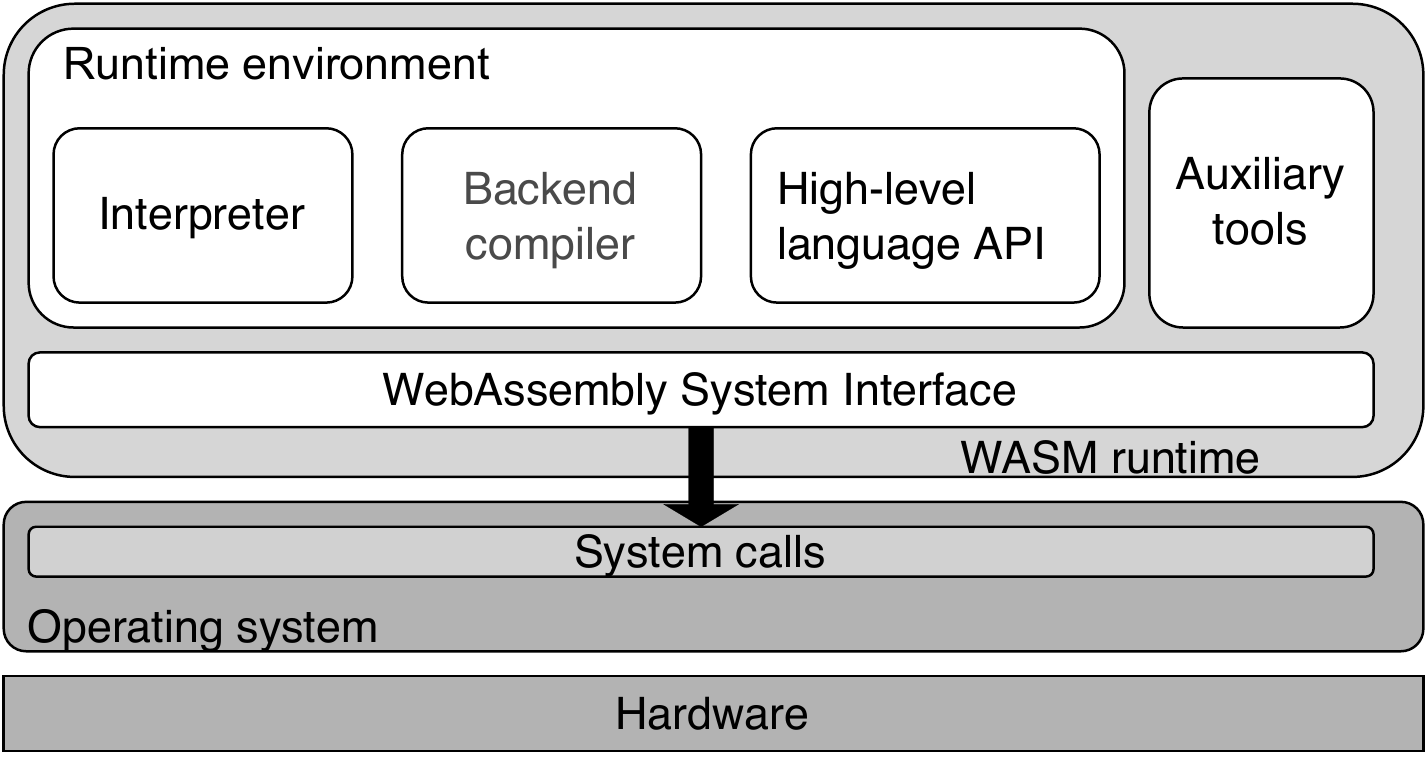}}
    \caption{The general architecture of a WASM runtime.}
    \label{figure:wasm_runtime_architecture}
\end{figure}

Based on the implementation of well known WASM runtimes ~\cite{wasm_runtime_architecture_link, wasmer, wasmtime, WAMR, wasm3, WASM-Edge, wasilink}, we have summarized the general architecture of WASM runtimes in Figure~\ref{figure:wasm_runtime_architecture}, which can be divided into six major components.

\textbf{Backend compiler.} WASM runtimes support executing WASM binaries in the following modes: interpreter mode, Ahead-of-Time compilation mode (AoT), and Just-in-Time compilation mode (JIT). WASM runtimes support compiling WASM binaries into native code before executing it locally using AoT compilers. To speed up the execution efficiency, some WASM runtimes use the just-in-time compilation of hot code through JIT compilers. JIT compilers and AoT compilers are considered backend compilers in the WebAssembly workflow.

\textbf{Interpreter.} Some WASM runtimes provide interpretive execution on the WASM binaries.

\textbf{Runtime environment.} The runtime environment supports allocating memory, performing stack operations, reporting execution error messages, and other features.

\textbf{High-level language API.} The WASM runtimes can be embedded in different high-level languages (e.g., C/C++, Java, Python, Rust, etc.) as a library to allow users to use WASM in any scenarios with various languages.

\textbf{WebAssembly system interface.} WASM runtimes provide WASM applications with WebAssembly system interface (WASI) \cite{wasidoc} as a modular system interface \cite{wasilink}, focusing on security and portability. WASI is the bridge between the sandbox environment and operating systems. WASI is an API that provides access to several OS-like features, including file operation and clock.

\textbf{Auxiliary tools.} WASM runtimes also provide handy little tools for the users, such as WASM module cache, WASM textual file format validation, etc.

\section{Characterization Methodology}\label{sec:methodology}
We first perform an empirical study to characterize WASM runtime bugs. Specifically, we seek to investigate: 1) \textbf{the taxonomy of bugs}, i.e., the reasons leading to the bugs, and 2) \textbf{the fix strategies}, i.e., how to address these bugs.

\begin{figure}[htb]
	\centerline{\includegraphics[width=0.75\textwidth]{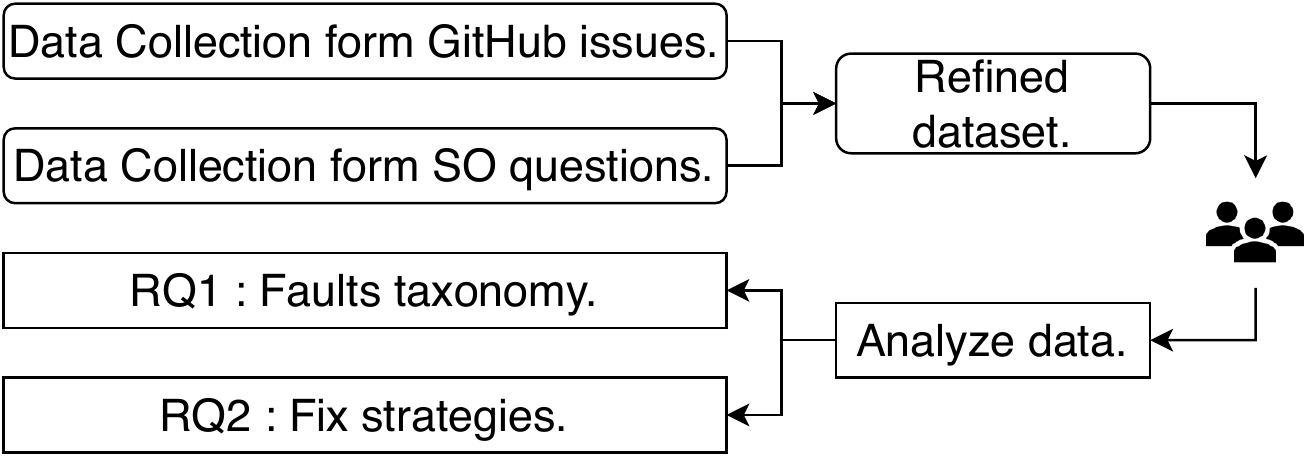}}
	\caption{Overview of the methodology.}
	\label{figure:methodology}
\end{figure}

To approach the answer, we collect and analyze the bug reports posted on Github and Stack Overflow, following the traditional empirical methods in the SE community \cite{empiricalstudy_wasmcompiler, ossbug, empiricalstudy_wasmbinaries, empirical_DLdeployment, empiricalstudy_DLapplications, empiricalstudy_TFbugs, qualitative_methods, pythonbug, platformbug, gccbug}. Figure~\ref{figure:methodology} shows the overview of our study methodology.

\subsection{Collection of WASM Runtime Bugs}

\subsubsection{Selecting WASM Runtimes}
As shown in Table~\ref{source_data}, we select three most popular WASM runtimes as target, including \texttt{wasmer} \cite{wasmer}, \texttt{wasmtime} \cite{wasmtime}, and \texttt{wasm-micro-runtime (WAMR)} \cite{WAMR}. We believe they are the most representative WASM runtimes for us to characterize real-world WASM runtime bugs across different implementations, as 1) all of them are mature projects (with over 100,000 LOC) that have gained the thousands of stars on GitHub, 2) they have covered different kinds of execution modes (i.e., Interpreter, JIT and AoT), and 3) they have implemented in different languages (i.e., Rust and C/C++).

\subsubsection{Data Collection from GitHub}

Following previous work \cite{empirical_DLdeployment, empiricalstudy_wasmcompiler, empiricalstudy_TFbugs, pythonbug, platformbug, gccbug}, we extract issues in the official GitHub repositories of the selected WASM runtimes. GitHub issues contain many bug information, including source code, detailed reports, and contributors’ discussions \cite{real-world_numerical_bug}. These characteristics make GitHub issues suitable for analyzing bug root causes and summarizing fix strategies. 
For details, we use the GitHub search API \cite{githubsearchapi} to extract the related issues on May 14, 2022. GitHub issues include various topics, including bug reports, feature requests, documentation updates, etc. Thus, to highlight the purposes of bugs, we take advantage of the bug issue label to identify related issues. 

We collect issues related to \texttt{wasmer} and \texttt{wasmtime} by filtering labels with ``bug''. Due to all the issues from \texttt{WAMR} are not labeled, we extract all the issues from \texttt{WAMR} for further analysis. Overall, we obtained 403 issues from \texttt{wasmer}, 167 issues from \texttt{wasmtime}, and 333 from \texttt{WAMR}.

\subsubsection{Data Collection from SO}
Initially, we also considered posts from Stack Overflow.

Each SO question has at least one tag based on its topics. We extract the posts related to the selected WASM runtimes on May 14, 2022. As a result, we obtain 41 posts for \texttt{wasmer}, 52 posts for \texttt{wasmtime}, and 4 posts for \texttt{WAMR}. Table~\ref{source_data} shows the collected raw data.

\begin{table}[t]
\caption{Statistics of our harvested dataset.}
\vspace{-0.1in}

\begin{center}
\label{source_data}

\setlength{\tabcolsep}{1mm}{\begin{tabular}{cccccc}
\toprule 
\textbf{Runtime} &\textbf{Stars} &\textbf{Commits} &\textbf{GitHub issues} &\textbf{SO posts} &\textbf{Total}\\
\midrule  
wasmer & 12,026 & 11,332 & 403 (179) & 41 (0) & 444 (179)\\
wasmtime & 7,360 & 9,754 & 167 (94) & 52 (0) & 219 (94)\\
WAMR & 2,720 & 686 & 333 (38) & 4 (0) & 337 (38)\\
\midrule 
\multicolumn{3}{c}{\textbf{Total}} &\textbf{903 (311)}&\textbf{97 (0)}&\textbf{1000 (311)}\\
\bottomrule 
\multicolumn{6}{l}{$^{\mathrm{*}}$The refined numbers are in the parentheses.}
\end{tabular}}
\vspace{-0.1in}
\end{center}
\end{table}

\subsubsection{Refining the Dataset}
We perform manual investigation on the collected data.
First, we filtered out issues and posts with no definite answers, to ensure the accuracy and certainty of bugs and fix strategies. Second, we exclude installation/build bugs, documentation bugs, and other issues and posts unrelated to WASM binaries' execution from the source data. 
Finally, as shown in Table~\ref{source_data}, the total number of WASM runtime bugs is 311. The scale of this dataset is comparable and more extensive than those used in existing bug-related studies \cite{empiricalstudy_wasmcompiler,empirical_DLdeployment, empiricalstudy_DLapplications, empiricalstudy_TFbugs, real-world_numerical_bug, auto_classify_SO, software_doc_issues} that also require manual inspection. All the 311 issues are from GitHub because all the reports we collected from SO are not related to the WASM binaries execution in WASM runtimes. It is probably because there are few WASM experts on SO since WASM is an emerging language. Therefore, WASM developers tend to report the bugs they encounter to the official WASM runtime repositories to seek immediate help.

\subsection{Labelling Bugs and Fix Strategies}
The refined 311 bug reports are used for distilling features and fix strategies through manual labelling by two authors and an intercessor.

\subsubsection{Pilot Labelling}
First, we randomly sample 50\% of the posts ($N=155$) from the selected WASM runtimes for pilot labeling. The first two authors of the paper jointly participate in the process. According to the WASM runtime architecture and the root causes, they create the bug categories and fix strategies by analyzing the GitHub issues.

\subsubsection{Reliability Analysis}

For reliability analysis, the first two authors independently label the remaining 40\% issues based on the taxonomy constructed in the prior stage. In detail, they mark each issue with the posted bug, fix strategy categories, and the issues that cannot be classified into the current taxonomies as a new category. To measure the reliability during the independent labelling, we employ the widely used Cohen’s Kappa indicator ($\kappa$) for bug and fix strategies of 0.921 and 0.915, indicating almost perfect agreement \cite{kappa_journal}. The agreement levels demonstrate the reliability of our labelling.

The divergence in the labelling process is then discussed and settled after the labeling process. For the newly added categories by the first two authors, we discuss them with the intercessor. As a result, we add two new categories to the bug taxonomy and three new categories into the fix strategy taxonomy. 
Furthermore, the first two authors independently label the remaining 10\% issues. During this process, no more bug taxonomy or fix strategy is added, indicating \textbf{saturation of the taxonomy}. After finishing the whole labelling stage, the Cohen’s Kappa indicator ($\kappa$) for bug and fix strategies is 0.929 and 0.925, showing almost perfect agreement \cite{kappa_journal}.
Additionally, the three authors involved in the taxonomy check the final labeling result together.

We will detail the bugs and fix patterns in the following sections.

\section{Taxonomy of WASM Runtime Bugs}\label{sec:RQ1}
\begin{figure*}[htb]
	\includegraphics[width=0.99\textwidth]{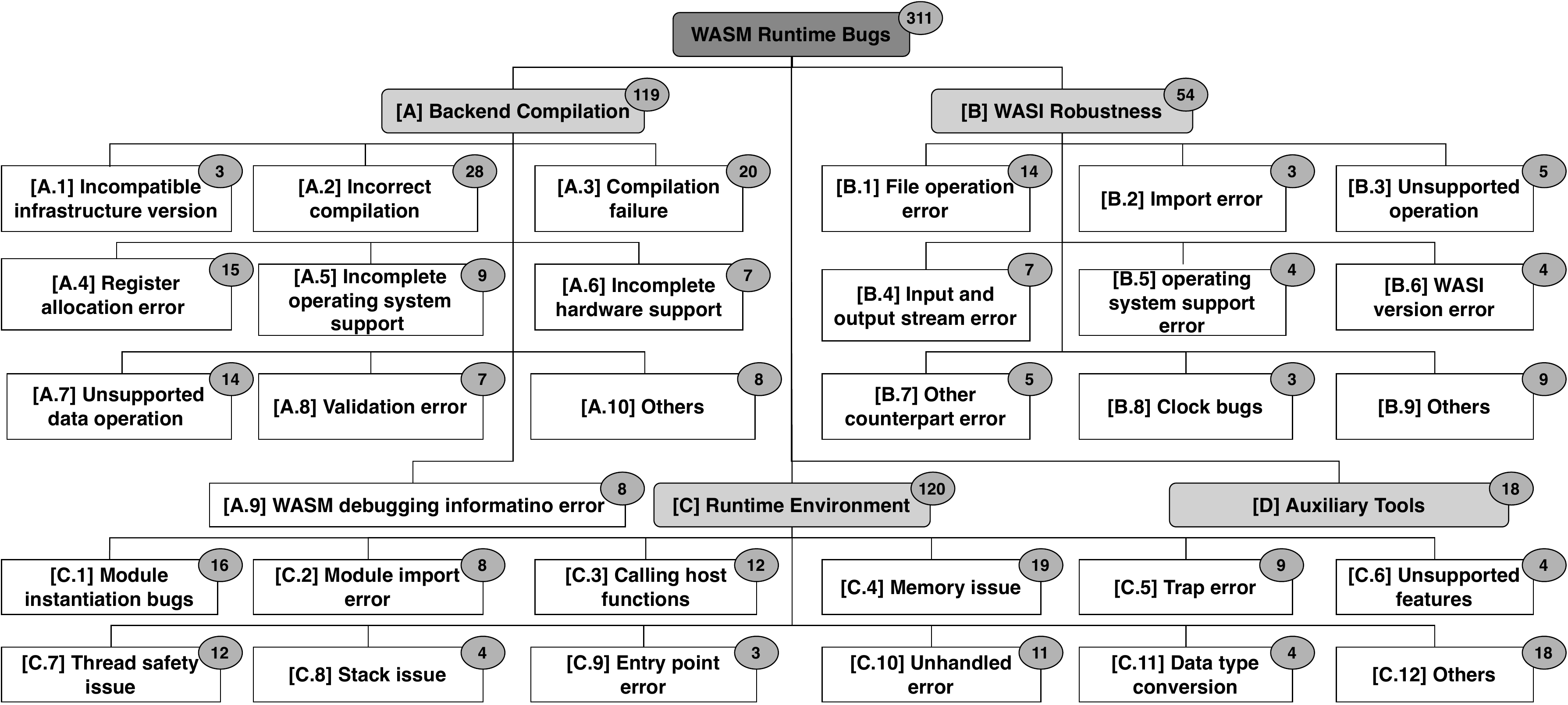}
	\caption{Taxonomy of bug symptoms. The number in the top right corner indicates the number of bugs for each category.}
	\label{figure:bug_taxonomy}
\end{figure*}

We present the hierarchical taxonomy of WASM runtime bugs according to the WASM runtime architecture (see \textbf{\ref{sec:background}}). As shown in Figure~\ref{figure:bug_taxonomy}, the taxonomy is organized into three-level categories, including a root category (\textit{WASM Runtime Bugs}), four inner categories linked to different components in a WASM runtime (e.g., \textit{Backend Compilation}), and 31 specific leaf categories (e.g., \textit{Register allocation error}).

The backend compilers (JIT compilers and AoT compilers) of the architecture are summarized into one inner bug category, called \textit{Backend Compilation (A)}, which converts WASM binaries into native code. The bugs in the lowest part in a WASM runtime are called \textit{WASI Robustness (B)}. The handy little tools in WASM runtimes are called \textit{Auxiliary Tools (D)}. Other bugs that occur while running WASM binaries are classified into \textit{Runtime Environment (C)}, including memory allocation, calling host functions, and so on. It is worth mentioning that bugs that occur while using high-level language API are either divided into \textit{Backend Compilation (A)} or \textit{Runtime Environment (C)}. WASM users could use the API to compile WASM binaries and take advantage of functionalities in the Runtime Environment, and it is an interface for users to make good use of a WASM runtime. Moreover, the interpreter part is merged in a leaf category of \textit{Backend Compilation (A)}, as only WAMR provides an interpreter, and only one bug is found in the interpreter.

\begin{tcolorbox}
\textbf{Highlight 1:} \textit{We construct a taxonomy of 31 leaf bug symptom categories in WASM runtimes, indicating the root causes and the diversity}. 
\end{tcolorbox}

\subsection{Backend Compilation} 
As the first stage of executing WASM binaries, backend compilation is used to translate WASM binaries into native code. 

In general, backend compilers convert WASM binaries into their intermediate representation (IR), allocate registers and optimize the code. Note that backend compilers could convert WASM binaries into the IR proposed in other compilation framework infrastructures (e.g., LLVM). The whole process needs to support various OSes and CPU architectures. We observe 119 bugs in this category, accounting for 38.3\% of all the classified bugs and covering 10 leaf categories.

Various backend compilers use their own IR as the intermediate step to translate WASM instructions to native code. During the process of compiling, compilers could generate incorrect IR or incorrect native code during the translation of WASM binaries. Besides, optimizing the code could also lead to an error. These bugs are summarized as \textit{Incorrect compilation (A.2)}. A compiler may raise an exception when generating native code or even \textit{fail to generate the native node (A.3)}, which accounts for 16.8\% bugs in \textit{Backend Compilation (A)}.
Moreover, some WASM runtimes rely on the existing compilation framework, such as LLVM. Thus, \textit{Using the incorrect version of infrastructure (A.1)} could lead to unexpected results, accounting for 2.5\% bugs in \textit{Backend Compilation (A)}.

\begin{figure}[htb]
	\includegraphics[width=0.75\textwidth]{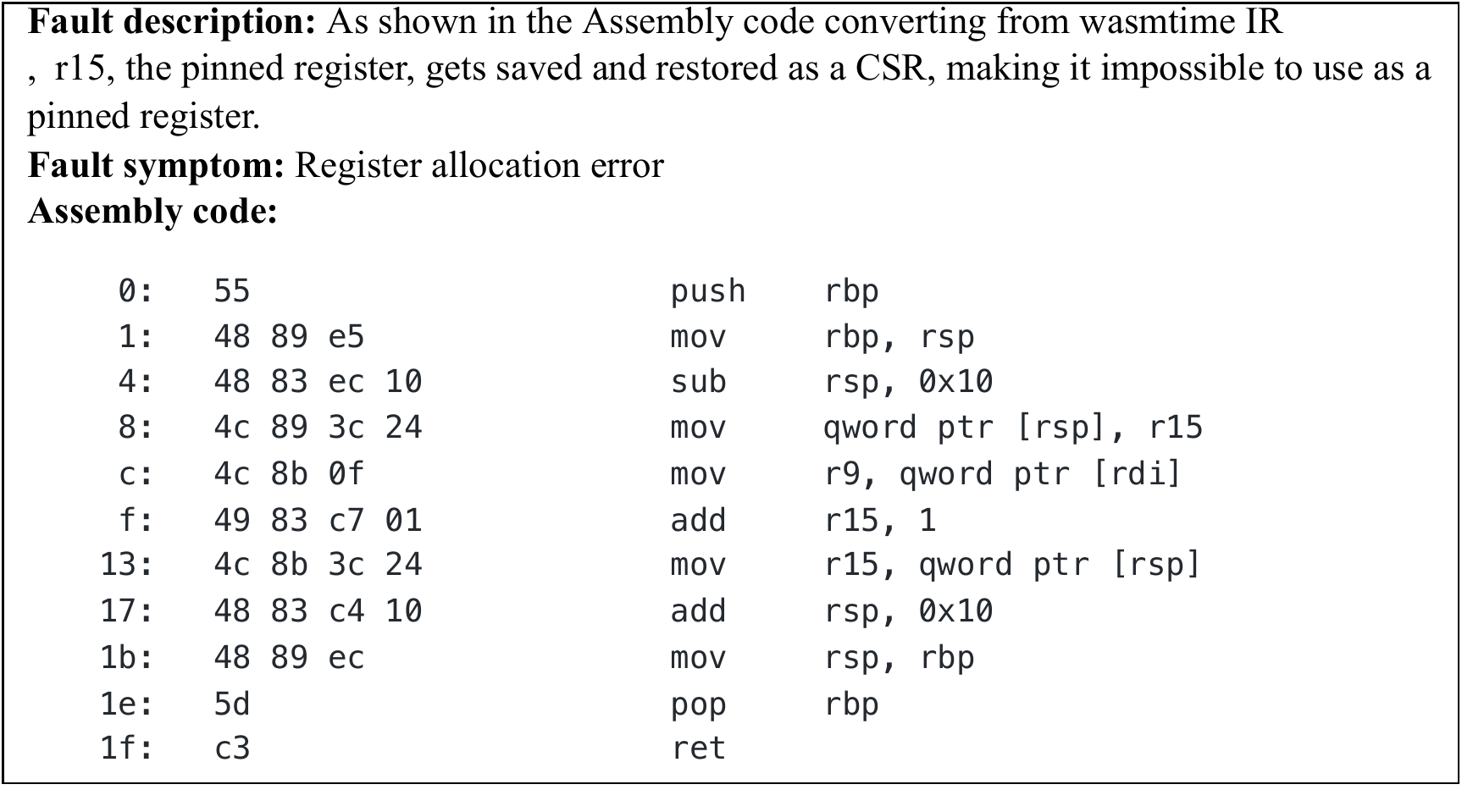}
	\caption{Example (a) - GitHub \texttt{wasmtime} issue \#4170}
	\label{figure:exampleA}
\end{figure}

\begin{figure}[htb]
	\includegraphics[width=0.75\textwidth]{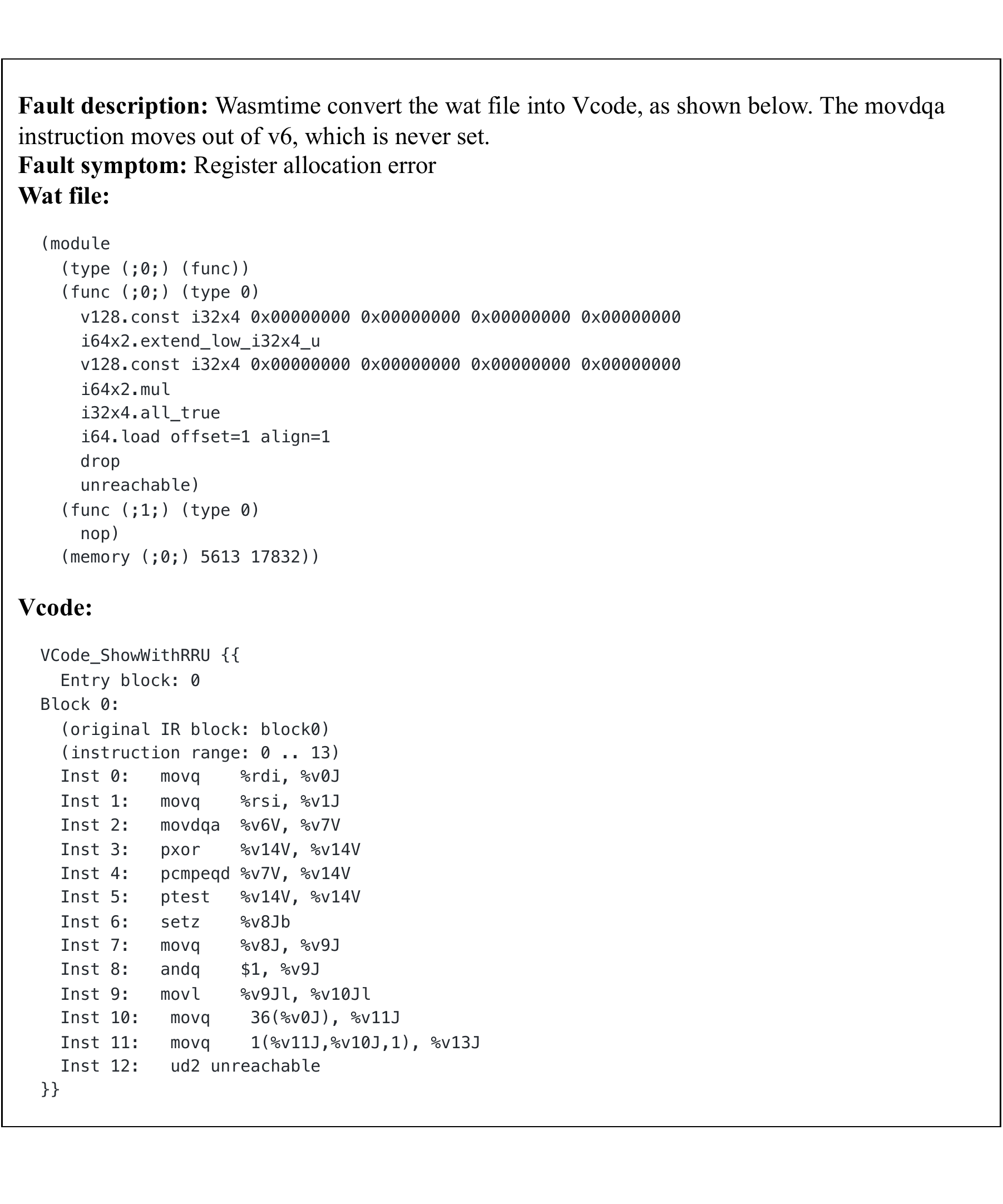}
	\caption{Example (b) - GitHub \texttt{wasmtime} issue \#3337}
	\label{figure:exampleB2}
\end{figure}

Besides converting WASM instructions into native machine instructions, the backend compilers must allocate registers. However, they may result in the \textit{Incorrect register allocation (A.4)}, including incorrectly using special registers, loading data from an unexpected register and exhausting registers. These bugs account for 7.6\% of bugs in \textit{Backend Compilation (A)}. As shown in Example (a) (Figure~\ref{figure:exampleA}), the backend compiler in \texttt{wasmtime} gets saved and restored in \texttt{r15} as a \texttt{CSR} (control and status register), which is expected to be used as a pinned register. The allocation of \texttt{r15} poses a bug. As another example in Example b) (Figure~\ref{figure:exampleB2}), \texttt{wasmtime} allocates registers for the given wat file. However, during lowering SIMD instructions, the allocation of registers shows a bug. The \texttt{movdqa} instruction moves out of \texttt{v6}, but \texttt{v6} is never set. This kind of bugs will cause panic during the execution of WASM binaries, and the execution process cannot be completed.
Most WASM runtimes only support JIT or AoT compilation, while \texttt{WAMR} also provides an interpreter to deal with WASM. There is only one bug in the interpreter. The interpreter could not correctly pass parameters to submodules, leading to an incorrect result. Moreover, this is summarized into \textit{Others (A.10)}.

In the compilation process, WASM runtimes run the WASM file across various \textit{operating systems (A.5)}. They account for 7.6\% of bugs in the current inner category. The backend compilers encounter problems only caused by specific operating systems and lack consideration for their particular circumstances. With its architecture and instruction set, WASM runtimes also run the WASM files across different CPUs. Some problems are only present in specific CPUs or specific architecture machines. These problems are summarized as \textit{Incomplete hardware support (A.6)} which account for 5.9\% bugs in Backend Compilation.

Further, we also observe that a portion (i.e., 11.8\%) of bugs in \textit{Unsupported data operation (A.7)}. For example, in the IR used by the backend compiler in \texttt{wasmtime}, the \texttt{srem.8} and \texttt{srem.16} are not supported. Besides, wasmtime converts the WASM binaries into cranelift IR before execution. It lacks the design of supporting the data operation in big endianness machines (e.g., GitHub \texttt{wasmtime} issue \#3288). Executing the clif file on s390 hardware shows wrong results only for i16, i32, and i64 types, while i8 passes these tests. The s390 architecture is big-endian, while the data operation in wasmtime was taken from the lower bites. Thus, the data operation of i16, i32, and i64 was not supported in the big-endianness machine. This kind of bug could pose different execution results on  or execution exceptions. This bug can result in inconsistent results or execution exceptions for the same WASM binaries executed on different machines. Besides, WASM specification introduced Single Instruction Multiple Data (SIMD) instructions to improve the execution efficiency. Backend compilers may lack the support for data operation related to SIMD instructions, such as the operation of the v128 data type.

During the compilation process, the verifier must validate the legalization of IR,  WASM instructions, and the temporary files emitted by AoT compilers. The \textit{incorrectness or strictness of validation (A.8)} causes errors in the backend compilation process, accounting for 5.9\% bugs in \textit{Backend Compilation (A)}.
To make it easier for the users to utilize the WASM runtimes and learn about the code error, the WASM runtime developers should consider the debugging information during the backend compilation. The \textit{WASM debugging information (A.9)} bugs lead to several consequences, such as failing to provide debugging information or even influencing the compilation of WASM information, accounting for 6.7\% bugs in \textit{Backend Compilation (A)}.

\begin{tcolorbox}
\textbf{Highlight 2:} \textit{Bugs in Backend Compilation account for 38.3\% of WASM runtime bugs, covering 10 leaf categories. A large proportion (23.5\%) of these bugs are thrown with incorrect compilation.}
\end{tcolorbox}

\subsection{WASI Robustness} 
WASI is the fundamental part of a WASM runtime, allowing WASM to run outside the web. WASI supports WASM with several OS-like features, including files, sockets and the clock. 

Each WASM runtime could implement its specific features. Bugs in this part account for 17.4\% of our dataset.

WASI allows the WASM binaries to perform file operations, including making a new directory, writing and reading files, deleting files, and so on. The most common bug related to WASI is \textit{File operation error (B.1)}, which accounts for 27.8\%. For example, users failed to rename a file through WASI by applying wasmer in the \texttt{wasmer} issue \#2297.
Besides, \textit{Input and output stream error (B.4)} and \textit{Clock Bugs (B.8)} are also found in this part, accounting for 13.0\% and 5.6\% of bugs in WASI Robustness individually. 

Different WASM runtimes implement their own WASI, which may lack the \textit{support for some operations (B.3)}. 9.3\% of bugs in WASI Robustness are triggered due to unsupported operations.

In addition to the basic functionalities, WASI relies on different WASI modules and versions. Frontend compilers convert the high-level language into WASM binaries which may include a few WASI modules and different versions. WASM should \textit{import different WASI modules (B.2)} to support specific functionalities. These imports encounter a few bugs, such as module not found, incompatible version, etc. These bugs account for 5.6\% in WASI Robustness. Due to the updated versions of runtimes, new WASI versions are continuously provided by the runtimes. Using the \textit{incompatible WASI version (B.6)} in WASM runtimes could be unsupported and account for 7.4\% of bugs in WASI Robustness.

Moreover, WASI is the bridge between WASM and the OS, which should support different OSes. The diverse in these OSes can result in \textit{Operating system support error (B.5)}, accounting for 5.6\% bugs in this category.

Furthermore, all the WASM runtimes should support WASI to interact with the low-level system, and \texttt{wasmer} also provides another application binary interface (ABI) to do this.

Bugs in this part are regarded as \textit{Other counterpart error (B.7)} that account for 9.3\% of bugs in this category.

\begin{tcolorbox}
\textbf{Highlight 3:} 17.4\% of bugs are related to WASI implementation, covering 9 symptom categories. In particular, 46.3\% of the bugs in this category are related to the basic functionalities of WASI (i.e., B.1, B.2, and B.4).
\end{tcolorbox}

\subsection{Runtime Environment}
After compiling WASM to native code, WASM runtimes support the WASM with an execution environment. 
The runtime environment supports WASM with module import, trap message tracking, metering computing cost, and other functionalities.
Bugs happen in the \textit{Runtime Environment (C)} account for 38.6\% of bugs in total.

We observe a significant proportion (20\%) of bugs about module operation in Runtime Environment, including \textit{Module instantiation bugs (C.1)} and \textit{Module import error (C.2)}. Specifically, 13.3\% of bugs in this category are related to WASM module instantiation. WASM module has to be instantiated before execution. These bugs are related to the instance allocator, module loading, multiple instantiation error, etc. High-level language API makes it easier for WASM developers to utilize WASM runtimes. The API could import modules from the host environment or other WASM modules. These bugs are about unknown imports, calling host functions, etc.

Functions in WASM could \textit{call the host functions defined in high-level language (C.3)} and account for 10\% bugs in \textit{Runtime Environment (C)}. This process contains bugs of parameter passing, finding host functions, etc.

\textit{Memory issue (C.4)} is a common kind of bugs when executing the WASM binaries, accounting for 15.8\% bugs in this category. These bugs are about memory management, including memory allocation, multi-memory support, out-of-memory error, memory release and memory growth.

When executing WASM binaries, WASM runtime could encounter bugs in dealing with \textit{traps (C.5)} and lead to an abortion, accounting for a total of 7.5\% of bugs in this category. These bugs are related to the process of the \texttt{unreachable} instructions in WASM binaries.
Besides, WASM runtimes sometimes do nothing with the errors, and the errors were not carefully reported to users. The WASM runtime should \textit{generate an exception or return a well-defined error (C.10)}.

In executing native code generated from WASM, many users encounter \textit{Thread safety issue (C.7)} and \textit{Stack issue (C.8)}. \textit{Thread safety issue (C.7)} refers to the thread safety when executing WASM binaries. \textit{Stack issue (C.8)} refers to the bugs about the stack, such as match rules for popping when calling WASM functions. These bugs account for a total of 13.3\% of bugs in \textit{Runtime Environment (C)}.

We also observe three \textit{bugs about the entry point of a WASM module (C.9)}.

The functions named ``\_main'', ``\_start'', ``main'', and ``start'' are regarded as entry points of a WASM module. A WASM runtime will call the entry point function by default without setting the function name through the command line or high-level language API. However, some WASM runtimes require each WASM module to hold an entry point which is too strict. These bugs are summarized as \textit{Entry point error (C.9)}.

Furthermore, when developers use high-level language API to do some operations of a WASM module, they may encounter \textit{data type conversion problems (C.11)}, accounting for 3.3\% of bugs in the current inner bug category.

Besides, the Runtime Environment can not meet all expectations of functionalities from users. Runtime Environment \textit{lacks support for some features (C.6)} that users need, which account for 3.3\% of bugs on its own.

\begin{tcolorbox}
\textbf{Highlight 4:} Most (i.e., 38.6\%) bugs occur in the Runtime Environment, covering a broad spectrum of symptoms (i.e., 12 leaf categories). Among them, memory issue is the most common, accounting for 15.8\% of bugs in this category. 
\end{tcolorbox}

\subsection{Auxiliary tools}
Besides executing WASM files, WASM runtimes provide users with handy little tools related to WASM, including validating the format of WASM files, WASM module cache, Wat and WASM files conversion and package manager. As different WASM runtimes differ significantly in this respect, this category is not classified into leaf categories. This category accounts for 5.8\% of all the classified bugs.
For example, in wasmer issue 2028, when passing environment variables into wasmer run via the \textit{--env} flag, the program will fail if the environment variable contains an `=,' which should be allowed. Moreover, when validating the format of WASM binaries, wasmer uses command \textit{wasmer validate} to do this. Although setting the parameter \textit{--enable-simd}, it incorrectly reports an error when validating a WASM module with SIMD.

\section{Fix Strategies of WASM Runtime Bugs}\label{sec:RQ2}
To figure out how developers fix various types of bugs, we distill their fix strategies in this section for each inner bug category. 
Due to bugs in categories \textit{Auxiliary tools} are either too specific or irrelevant to WASM runtime themselves, and they only account for 5.8\% of bugs, we do not study the fix strategies for them. 
We have summarized the general fix strategies for the remaining three inner symptom categories. As shown in Figure~\ref{figure:fixA}, ~\ref{figure:fixB} and ~\ref{figure:fixC}, the X axis shows each leaf bug category in Figure~\ref{figure:bug_taxonomy}, and the Y axis represents the corresponding fix strategies following with their totally used frequency under the inner category. We elaborate on the summarized fix strategies of their frequent symptoms and demonstrate some examples of bugs and corresponding fixes in the real world.

\subsection{Fix Strategies for Backend Compilation} 

\begin{figure}[htb]
	\includegraphics[width=0.85\textwidth]{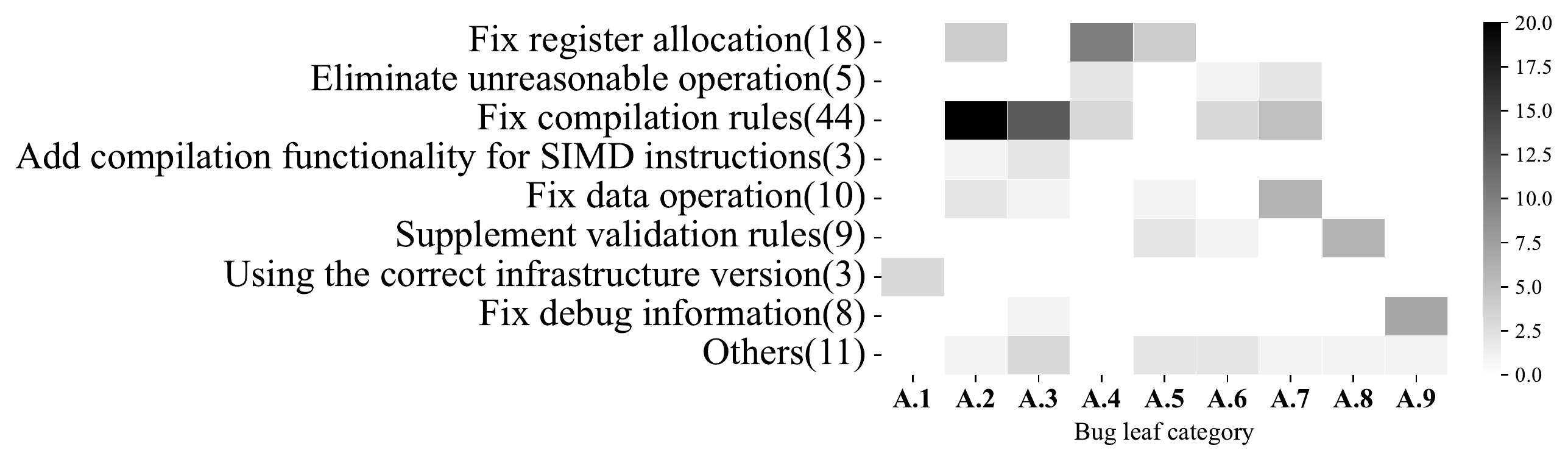}
	\caption{Distribution of fix strategies for Backend Compilation.}
	\label{figure:fixA}
\end{figure}

We summarize eight systematic fix strategies for bugs in Backend Compilation and illustrate the distribution of these strategies on leaf categories in Figure \ref{figure:fixA}.

\textbf{Fix compilation rules.} 39.6\% of bugs in Backend Compilation can be solved by modifying compilation rules in different backend compilers. Compilation rules are the guide for translating WASM instructions to native code. For example, \texttt{wasmtime} developers modify the \texttt{inst.isle} file to change the rules of emitting native code. \texttt{Wasmer} fixes the emitter file for different CPU architectures to emit native code.
This fix strategy covers five bug symptoms and is especially frequently adopted in the \textit{Incorrect compilation (A.2)} and \textit{Compilation failure (A.3)} bug categories. For example, 71.4\% of \textit{Incorrect compilation (A.2)} bugs are fixed after modifying the compilation rules in the backend compilers to support more reasonable translation. As for \textit{Compilation failure (A.3} bugs, the backend compilers may encounter unexpected exceptions and abortion due to the unreasonable compilation rules. Therefore, developers fix these bugs by changing the compilation rules to meet the actual requirements and support emitting correct native code in most cases.

\begin{figure}[htb]
	\includegraphics[width=0.75\textwidth]{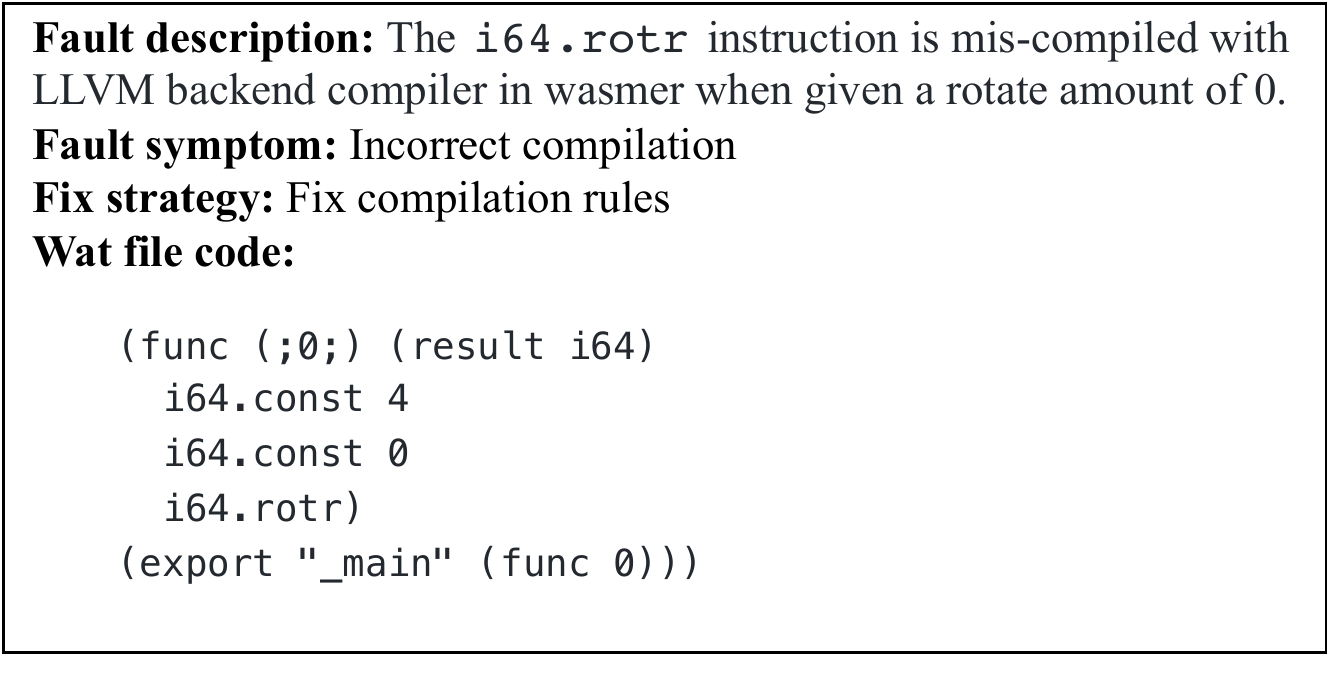}
	\caption{Example (c) - GitHub \texttt{wasmer} issue \#2215}
	\label{figure:exampleB}
\end{figure}

As shown in Example (c) (Figure ~\ref{figure:exampleB}), a developer reports that the \textit{i64\.rotr} instruction in WASM is incorrectly compiled with LLVM in \texttt{wasmer} when given a rotate amount of 0. The corresponding fix is to modify the translation process of this instruction in the LLVM backend compiler in the \texttt{wasmer}.
Compilation rules such as lowering rules guide translating WASM instructions to native code. 

Even worse than emitting incorrect native code is that the backend compilers fail to compile some instructions or the whole WASM module. For example (\texttt{wasmtime} issue \#2347), a user reports that the backend compiler in \texttt{wasmtime} (V0.20 and main branch) fails to compile a WASM module. The \texttt{wasmtime} developers fix it by modifying the compilation rules. In detail, they do block manipulation in the \texttt{wasmtime} translation of some table-related instructions and explicitly call the \textit{ensure\_inserted\_block()}.

\textbf{Fix register allocation.} 16.2\% of bugs in Backend Compilation, involving three frequent bug categories, can be fixed by changing the register allocation. As aforementioned in Section~\ref{sec:RQ1}, while generating native code from WASM instructions, the backend compilers are expected to allocate registers. Nevertheless, they may use incorrect registers, load data from an unexpected register, exhaust registers, etc. Various WASM instructions and instruction set architectures (ISA) make it hard for the backend compilers to allocate suitable registers.

\textbf{Fix data operation.} The fix strategy is used for 9.0\% of bugs in Backend Compilation, covering a wide range of bug categories, including \textit{Incorrect compilation (A.2)}, \textit{Compilation failure(A.3)}, \textit{Incomplete operating system support (A.5)} and \textit{Unsupported data operation (A.7)}. The strategies include fixing data alignment, adding support for i8, i16, fixing byte order, dealing with undefined upper bits, converting data types, returning multi-value data, supporting v128 data type, etc.

\textbf{Supplement validation rules.} Supplement rules of verifier will tackle the problems in validation, repairing 8.1\% of bugs in Backend Compilation, and mainly fix the \textit{Validation error (A.8)}. For example (\texttt{wasmer} issues \#2187), a developer reports that he can not get the memory page in WASM of 65536, although the user sets memory minimum and maximum sizes range 0..65536 inclusive by WASM instructions. The verifier is expected to block 65537 and higher. However, \texttt{wasmer} only works on 65535 and lower. This corresponding fix strategy is to modify its validation rules.

\textbf{Fix debug information.} This fix strategy repairs 7.2\% of bugs in \textit{Backend Compilation (A)}, dealing with 87.5\% bugs in \textit{WASM debugging information error (A.9)}.\textit{WASM debugging information error (A.9} may lead to the failure of providing incorrect debug information that misleads the users. By fixing debug information, these bugs could be well settled.

\textbf{Eliminate unreasonable operation.} Some functionalities in backend compilers of WASM runtimes lead to an unexpected consequence. These functionalities are meaningless and need to be limited. For example (wasmtime issues \#2883), wasmtime users try to use \texttt{ssub\_sat} with two I64 values. They use cranelift-object with a triple in the intermediate presentation (IR) in the wasmtime backend compiler. It is worth mentioning that \texttt{ssub\_sat} is a vector command, but it is used with a scalar. Moreover, the developers eliminate the unreasonable operation, limiting saturating arithmetic instructions: \texttt{uadd\_sat}, \texttt{sadd\_sat}, \texttt{usub\_sat}, and \texttt{ssub\_sat}, and applying them only to vector types. This kind of fix strategy fixes 4.5\% of bugs in Backend Compilation.

\textbf{Add compilation functionality for SIMD instructions.} Some WASM runtimes do not support the intact functionalities to deal with SIMD instructions. Adding the support will address the bugs. This strategy fixes 3 bugs in \textit{Incorrect compilation (A.2)} or \textit{Compilation failure (A.3)} related to SIMD instructions.

\textbf{Using the correct version of the infrastructure.} Since some backend compilers in WASM runtimes rely on existing frameworks such as LLVM, adjusting the LLVM's version can handle some problems. This fix strategy fixes all the bugs in \textit{Incompatible infrastructure version (A.1)}.

\begin{tcolorbox}
\textbf{Highlight 5:} We identify eight systematic fix strategies for bugs in Backend Compilation. The three most common strategies are fixing compilation rules, registering allocation, and data operation, resolving 39.6\%, 16.2\%, and 9.0\% of bugs in this category, respectively.
\end{tcolorbox}

\subsection{Fix Strategies for WASI Robustness} 

\begin{figure}[htb]
	\includegraphics[width=0.85\textwidth]{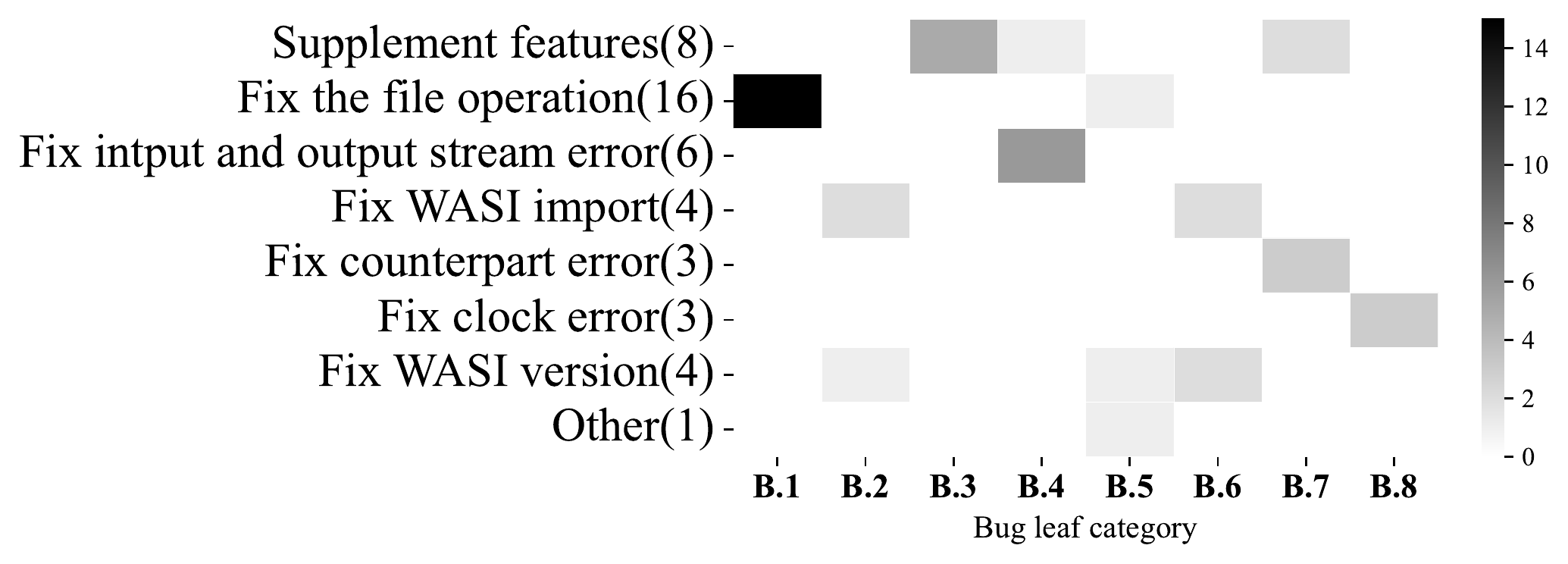}
	\caption{Distribution of fix strategies for WASI Robustness.}
	\label{figure:fixB}
\end{figure}

As illustrated in Figure \ref{figure:fixB}, we identify seven frequent fix strategies for bugs in WASI Robustness.

\textbf{Fix the file operation.} This strategy fixes 35.6\% bugs in WASI Robustness, including all the bugs in \textit{File operation error (B.1)} and half of the bugs in \textit{Operating system support error (B.5)}. For example (\texttt{wasmer} issue \#2297), a user reports that renaming a temporary file through a WASM file fails and prints \texttt{unable to rename temporary}. The developers fix the implementation of WASI to allow this operation.

\textbf{Supplement features.} 
WASM runtimes are expected to implement the necessary features.

However, some WASM runtimes do not fulfill this expectation. In such cases, WASM runtime developers need to implement the expected functionalities in WASI and thus the \textit{Unsupported operation (B.3)} bugs can be fixed.

\textbf{Fix input and output stream error.} When the required message is not successfully printed, or the necessary information is not correctly imported into WASM, the \textit{Input and output stream error (B.4)} occurs. In these cases, developers need to fix the input and output streams. This fix strategy fixes 13.3\% bugs in WASI Robustness.

\textbf{Fix WASI import \& Fix the WASI version.} The two strategies are mainly used to tackle the \textit{Import error (B.2)} and \textit{WASI version error (B.6)} which occur when using different modules from WASI. The developers fix WASI import to use the suitable WASI module to support specific functionalities, and fix WASI version to make the WASM binaries compatible with the current circumstances. The two strategies all fix 8.9\% of bugs in WASI Robustness.

\textbf{Fix counterpart error.} This fix strategy can resolve the \textit{Other counterpart error (B.7)}, accounting for 6.7\% of bugs in WASI Robustness, which could be regarded as the repair method for all the counterparts ABI.

\textbf{Fix clock error.} The fix strategy only focuses on the \textit{Clock bugs (B.8)} in WASI Robustness and reslove 6.7\% bugs.

\begin{tcolorbox}
\textbf{Highlight 6:} We distill seven systematic strategies for bugs in WASI Robustness. The most common one is fixing the file operation, which resolves 35.6\% of bugs in this category.
\end{tcolorbox}

\subsection{Fix Strategies for Runtime Environment.} 
\begin{figure}[htb]
	\includegraphics[width=0.85\textwidth]{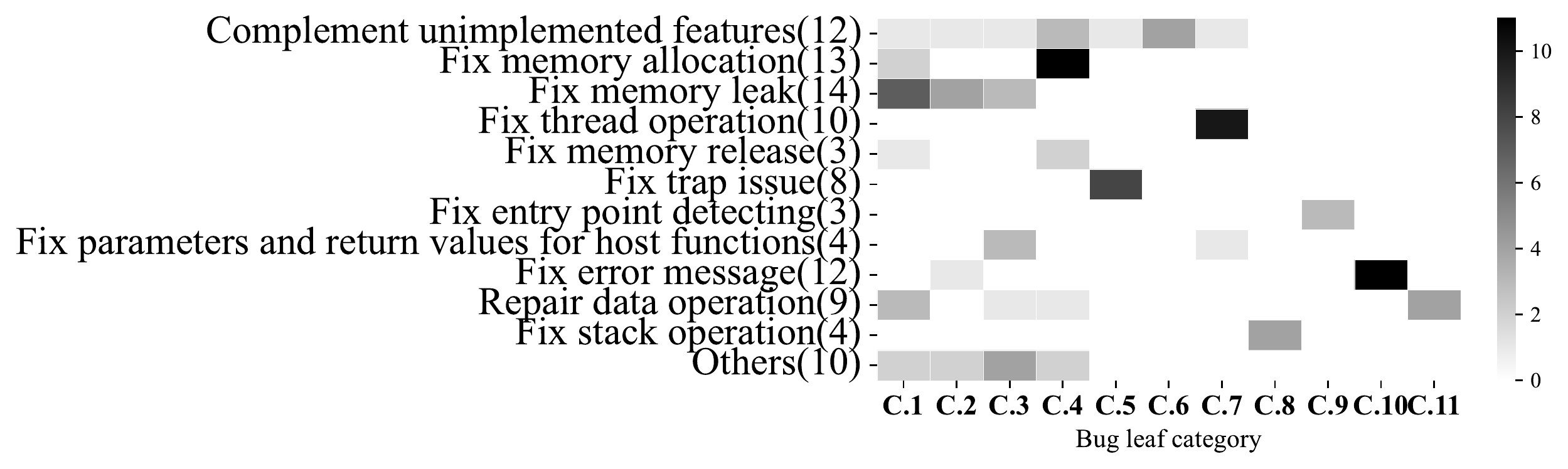}
	\caption{Distribution of fix strategies for Runtime Environment.}
	\label{figure:fixC}
\end{figure}

We identify eleven frequent fix strategies for bugs in Runtime Environment and Figure \ref{figure:fixC} shows the distribution.

\textbf{Fix memory allocation \& Fix memory leak \& Fix memory release.} 29.4\% of bugs in Runtime Environment can be resolved by the three fix strategies for memory management. The three strategies mainly fix the \textit{Module instantiation bugs (C.1)} and \textit{Memory issues (C.4)}. After compiling the WASM binaries into native code, WASM runtimes need to instantiate the current WASM module. Errors about instance allocation and module loading errors could happen. Besides, other bugs related to multi-memory support, out-of-memory, and memory growth could occur. The developers mainly use these methods to deal with such cases.

\textbf{Fix error message.} This fix strategy is to modify the error message to make it more reasonable or catch unexpected failures with an error message. This strategy fixes 11.8\% bugs in Runtime Environment. For example, a user reports that (\texttt{wasmer} issues \#868) the WASM runtime gives the wrong message about the error in the execution. In another case, the user reports that (\texttt{wasmer} issue \#830) the WASM runtime should throw an error instead of the occurrence of the panic when giving a string instead of a digit in the WASM program. Both two cases need to be fixed by modifying the error message.

\textbf{Complement unimplemented features.} The fix strategy resolves a wide range of bugs in the Runtime Environment, including Module instantiation bugs (C.1), Module import error (C.2), Calling host functions (C.3), Memory issue (C.4), Trap error (C.5), Unsupported features (C.6) and Thread safety issue (C.7), accounting for 11.8\% of the total number.

\textbf{Fix thread operation.} 83.3\% of the bugs in the Thread safety issue (C.7) are fixed by this resolution. For example (\texttt{WAMR} issue \#1144), a user reports that the \textit{wasm\_runtime\_atomic\_wait} is not thread-safe and one of subthreads call \textit{wasm\_runtime\_spawn\_exec\_env} return nullptr. The developers use this strategy to fix atomic wait to be thread-safe by a lock.

\textbf{Fix trap issue.} This strategy is entirely used to fix trap error (C.5), accounting for 7.8\% of bugs in Runtime Environment, including fixing trap catch, complementing the missing trap information, and so on.

\textbf{Repair data operation.} 8.9\% of bugs in Runtime Environment are fixed by repairing data operation, which mainly address the data type mismatch between WASM and high-level language API. Besides, it also fixes data alignment. All the Data type conversion (C.11) bugs are fixed by repairing data operation.

\textbf{Fix parameters and return values for host functions.} As suggested in Figure \ref{figure:fixC}, 25\% of bugs in Calling host functions (C.3) are fixed by modifying the parameters and return values because there is a mismatch or passing error between the parameters and return values.

\textbf{Fix entry point detecting.} 2.9\% of bugs in Runtime Environment are resolved by fixing the detection of the default entry point. The WASM runtime needs to check the existence of the specially named entry function before execution.

\textbf{Fix stack operation.} All the bugs in \textit{Stack issues (C.8)} are resolved by this strategy. The WASM runtime is expected to fix the order of the items when popped from the stack to match the WASM invocation rules.

\begin{tcolorbox}
\textbf{Highlight 7:}  The fix strategies for bugs in Runtime Environment are diverse. 

\texttt{WAMR} is the runtime with the best thread support of the three. Developers use \texttt{WAMR} to apply threads more often than \texttt{wasmer} and \texttt{wasmtime}, thus revealing more problems in \texttt{WAMR} of the thread safety bugs.
\end{tcolorbox}

\section{Pattern-based Bug Detector for WASM Runtimes}\label{sec:testsuite}

Our aforementioned analysis suggests that most bugs have specific patterns and share similarities across different WASM runtimes. Thus, in this section, we seek to develop a pattern-based bug detection framework to identify bugs in WASM runtimes. Our key idea is to construct test cases that can trigger various kinds of bugs we summarized in Section~\ref{sec:RQ1}. Specifically, we seek to construct one or more test cases to trigger each of the summarized bug category. Note that, the constructed test cases are either re-constructed from the bug reports or we create the from scratch according to the bug patterns.

Next, we will present the details of our bug-triggering test cases for the three major bug categories: \textit{Backend compilation (A)}, \textit{WASI Robustness (B)} and \textit{Runtime envrionment (C)}. Note that, \textit{D. Auxiliary Tools} is an additional part provided by WASM runtimes, and the tools provided by WASM runtimes in this part vary considerably. Thus, the category \textit{Auxiliary Tools (D)} is not considered in this section. Further, for some leaf categories, it is hard for us to construct test cases, which thus are not covered by our detection framework. In total, our detection framework is constituted of the following 19 bug detectors.

\subsection{Bug detectors for Backend Compilation}

\textit{[A.2] Incorrect compilation.} SIMD instructions are the newly introduced features for WASM binaries. WASM runtimes show the bug pattern when compiling specific SIMD instructions, such as using \texttt{i64x2} and \texttt{i32x4} to simulate the v128 type and the optimization for them. To identify such bugs, we select typical WASM binaries with these instructions to do the detection.

\textit{[A.3] Compilation failure.} Even worse, WASM runtimes could fail to generate native code for some instructions, especially those with v128 as parameters or the large WASM modules. We construct the bug detector to detect the \textit{select} with v128 as the parameters, a large WASM module such as Ti database with all the backend compilers in WASM runtimes.

\textit{[A.4] Register allocation error.} WASM runtimes could load data from an undefined register or get fused with other instructions when compiling the specific instructions, such as \textit{i64x2.extend\_low\_i32x4\_u} and \textit{f64x2.replace\_lane}. To identify such bugs, we select typical WASM binaries with these instructions to do the detection.

We extract the specific OS-related bug-triggering WASM modules for \textit{[A.5] Incomplete operating system support} and unsupported data operation such as alignment of SIMD for \textit{[A.7] Unsupported data operation}.

We use the max value linear memory to detect the \textit{[A.8] Validation error} and WASM binaries fromm bug reports which easily trigger debugging information to detect the \textit{[A.9] WASM debugging information error}.

\subsection{Bug detectors for WASI Robustness}
\textit{[B.1] File operation error.} Different WASM runtimes show similar bug patterns about file operation errors. These easily bug-triggering file operations include renaming, moving, counting, and mapping. Thus, based on the shared file operation bug types mentioned in the bug issues, we design the bug detector to detect these bug types. For example, we test whether WASM runtimes could rename a file or report error information when the file does not exist. Besides, the detector could test whether WASM runtimes can move a file, count the file number in a directory, or do a mapping operation.

\textit{[B.2] Import error.} The most commonly found bug about import is that some WASM runtimes could not support importing multiple WASI versions in one WASM module. Thus, we import both \textit{wasi\_snapshot\_preview1} and \textit{wasi\_unstable}, the most used WASI versions, in one WASM module to detect this bug.

We extract the WASM binaries, including the unsupported pre-opened directories with / and ./ for WASI, to detect the bug in \textit{[B.3] Unsupported operation}.

\textit{[B.4] Input and output stream error.} To support detecting bugs about standard input and output stream, we use C++ and compile the C++ program into WASM binaries by emscripten\cite{Emscripten} to see the rights and types in \textit{\_\_wasi\_filestat\_t}. If the WASM runtime could not successfully print the expected result, it could be a bug. For example, wasmtime print OS error when detecting this kind of bug, which the developer confirms.

\textit{[B.5] Operating system support error.} There are two OS-specific parts of WASI implementation: clocks and polling. Due to the difference among OSes, the same operation could fail in a specific OS. For example, the QuickJS engine based on WASM binaries only fails in windows due to the differences between POSIX and windows async APIs. We extract the QuickJS engine from bug issue to detect this bug. 

\subsection{Bug detectors for Runtime environment}
\textit{[C.1] Module instantiation faults.} WASM runtimea provide various high-level language APIs for users to execute WASM binaries embedded in different applications. When running WASM binaries in a high-level language, the first step is to load the WASM module from a file or directly load the textual format WASM module in a string variable. And then, instantiate the WASM module, including validating the WASM module, compiling the WASM binaries with the appointed backend compiler, allocating the memory allocation for the table, global, etc. However, WASM runtimes could not support the instantiation for an empty module. Some WASM runtimes will encounter memory leaks when instantiating multiple WASM modules in a short time. Thus, we use the bug detector to detect whether WASM runtimes support instantiates an empty WASM module and whether it will lead to memory leaks when instantiating multiple WASM modules in a short period.

\textit{[C.2] Module import error.} We observe that some WASM runtimes omit the step to check the index of imported items, such as skipping to report the error of `index out of bounds` errors when import\_global\_index is greater than imports. globals length. We extract the related WASM binaries from the raw bug report to detect this bug by the bug detector.

\textit{[C.3] Calling host functions.} We use the bugs detector for this kind of bug to detect whether WASM runtimes could support importing a self-defined module, not only from 'env.' Besides, some WASM runtimes show the bug pattern about mis-mapping multiple host functions. We use the bug detector to test whether WASM runtimes could successfully run the functions by importing them in the correct order or if the runtime could inspect the mapping by importing them in the wrong order and report the error message.

\textit{[C.4] Memory issue.} By the bug detector, we detect whether WASM runtimes could grow the linear memory dynamically. We extract the WASM module from bug issues and modify it to grow the memory using \textit{memory.grow} instruction and using \textit{memory.size} instruction to check the linear memory size after the growth.

\textit{[C.5] Trap error.} These bugs are related to the process of the \textit{unreachable}
instructions in WASM modules. By the bug detector, we use a WASM module with \textit{unreachable} instructions to test whether WASM runtimes could successfully break the execution and report the information in the location where \textit{unreachable} is.

\textit{[C.9] Entry point error.} WASM runtimes are expected to regard the function labeled with 'start' or '\_start' as the entry point and execute this function default and allow the WASM module without an entry point. WASM runtimes show a similar bug pattern about the entry point: do not run the entry point function or reject the WASM modules without an entry point. We construct the WASM module without or with an entry point to detect this kind of bug.

\textit{[C.10] Unhandled error.} Some WASM runtimes usually encounter panic directly without any operation to avoid it by reporting the error information. The most commonly found are unhandled errors with unsupported operation and invalid access to the data section. We extract typical WASM module examples to detect this bug.

\subsection{Reliability of the bug detectors}
As a portion of the bug detectors are curated by ourselves based on the code snippets and bug description provided in the bug reports, we first need to evaluate the reliability of the constructed bug detector. Note that, we already have the ground truth, i.e., WASM runtimes (with specific version) have some kinds of bugs. Thus, we make effort to reconstruct the environment (i.e., OS, WASM runtime version, configuration, etc.) to replicate the reported bug for each category. At last, the bug detector can trigger the reported bugs, which suggest the reliability of our detection framework.

\subsection{Detecting new bugs}
As the bug detector we create was constructed based on the knowledge summarized from wasmer, wasmtime, and WAMR, we further apply it to different WASM runtimes, seeking to identify new bugs. 

\textbf{Experimental Setting.}
In this experiment, besides the studied WASM runtimes (\texttt{wasmer}, \texttt{wasmtime} and \texttt{WAMR},), we further consider two unexplored runtimes, i.e., \texttt{wasm3} and \texttt{WASMEdge}, to investigate the generalizability of our study. 

The bug detector is applied to the following WASM runtimes: wasmer 2.3.0, wasmtime 0.38.0, WAMR 05-18-2022, wasm3 0.5.0, WASMEdge 0.9.1 on different execution modes (interpreter, AoT, JIT) and across three different operating systems (\texttt{macOS 10.15}, \texttt{Ubuntu 20.04}, and \texttt{Windows 11}).

\textbf{Result.}
As shown in Table~\ref{evaluation}, we find 53 new bugs, covering all the tested WASM runtimes. By the time of this submission, 14 of them have been confirmed by the developers, with 6 already been fixed in the main branch based on our suggestions. 

\begin{table}[t]

\caption{The experiment result of the bug detector. We mark a leaf category on a WASM runtime as \checkmark if it passes all the execution modes across all the OS platforms. Otherwise, it is marked with the number of detected bugs.}

\label{evaluation}

\begin{tabular}{| p{55mm}| p{10mm} | p{15mm} | p{10mm} | p{10mm} | p{15mm} | }

\hline
\textbf{Leaf category}&\textbf{wasmer}&\textbf{wasmtime}&\textbf{WAMR}&\textbf{wasm3}&\textbf{WasmEdge}\\

\hline

[A.2] Incorrect compilation & 1 & \checkmark & 3 & 1 & 2\\

[A.3] Compilation failure & 1 & 1 & 1 & 2 & 3\\

[A.4] Register allocation error & \checkmark & \checkmark & 1 & \checkmark & \checkmark\\

[A.5] Incomplete operating system support & \checkmark & \checkmark & 1 & \checkmark & 1\\

[A.7] Unsupported data operation & \checkmark & \checkmark & 1 & \checkmark & 1\\

[A.8] Validation error & \checkmark & \checkmark & 1 & 1 & \checkmark\\

[A.9] WASM debugging information error & 1 & \checkmark & 1 & \checkmark & 2\\

[B.1] File operation error & \checkmark & \checkmark & 3 & 4 & 2\\

[B.2] Import error & \checkmark & \checkmark & \checkmark & \checkmark & 1\\

[B.3] Unsupported operation & \checkmark & \checkmark & \checkmark & 1 & 1\\

[B.4] Input and output stream error & \checkmark & 1 & 1 & 1 & 1\\

[B.5] Operating system support error & \checkmark & \checkmark & \checkmark & \checkmark & \checkmark\\

[C.1] Module instantiation faults & \checkmark & \checkmark & \checkmark & \checkmark & 3\\

[C.2] Module import error & \checkmark & \checkmark & \checkmark & 1 & \checkmark\\

[C.3] Calling host functions & \checkmark & \checkmark & \checkmark & \checkmark & 1\\

[C.4] Memory issue & \checkmark & \checkmark & 1 & \checkmark & 1\\

[C.5] Trap error & \checkmark & \checkmark & \checkmark & 1 & \checkmark\\

[C.9] Entry point error & \checkmark & \checkmark & \checkmark & \checkmark & \checkmark\\

[C.10] Unhandled error & \checkmark & \checkmark & \checkmark & 2 & 1\\

\hline

\end{tabular}
\end{table}

\begin{figure}[t]
	\includegraphics[width=0.75\textwidth]{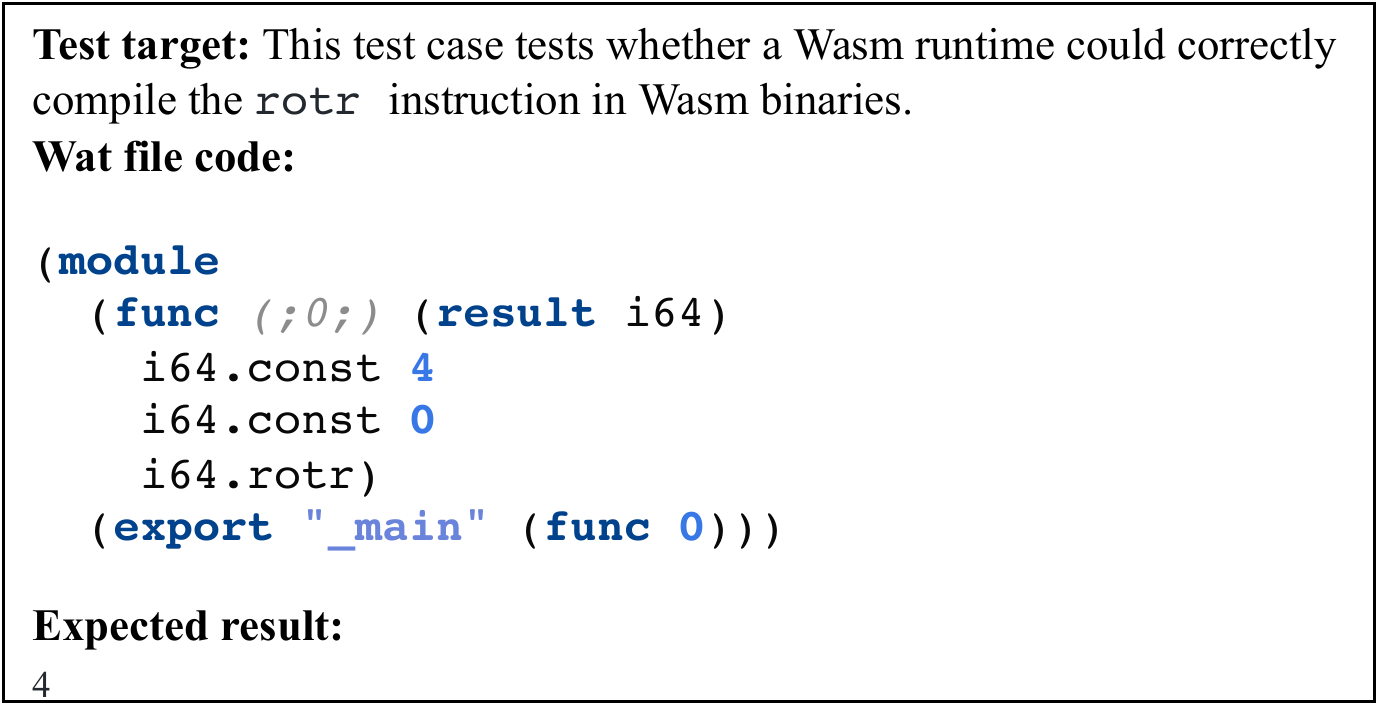}
	\caption{Case study of \texttt{[A.3] Incorrect compilation}}
 \vspace{-0.1in}
	\label{figure:exampleD}
\end{figure}

\textbf{Case Studies.}
As shown in Figure~\ref{figure:exampleD}, it is expected to print the number 4 when testing the \texttt{rotr} instruction for WASM binaries. However, the actual output in WAMR is a random number. Every time executing, it leads to a different output. The developers have confirmed it is a bug and fixed it in the main branch, dealing with the parameter 0 separately. This bug belongs to \textit{Incorrect compilation (A.3)}.

\begin{figure}[t]
	\includegraphics[width=0.75\textwidth]{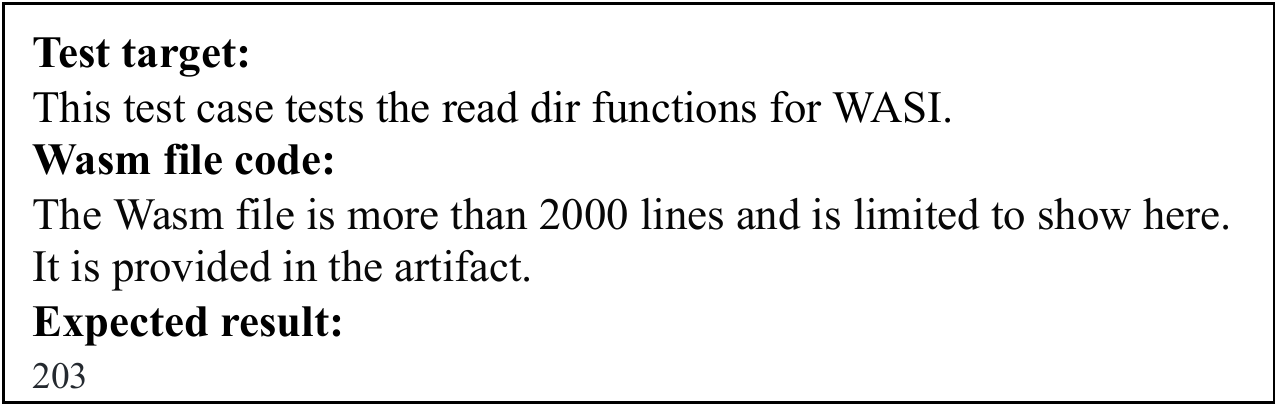}
	\caption{Case study of \texttt{[B.1] File operation error}}
 \vspace{-0.1in}
	\label{figure:exampleE}
\end{figure}

An additional example is shown in Figure~\ref{figure:exampleE}. It is expected to print the correct directory number 203 when testing \texttt{WASI} in the runtime. However, \texttt{WasmEdge} prints 147 as a result, which is already confirmed as a new bug by the developers. Once the number of files is larger than 147, it will be truncated in WasmEdge. 
And the file renaming belongs to \textit{[B.1] File operation error} fails in wasm3, which is also confirmed.

\begin{figure}[htb]
	\includegraphics[width=0.75\textwidth]{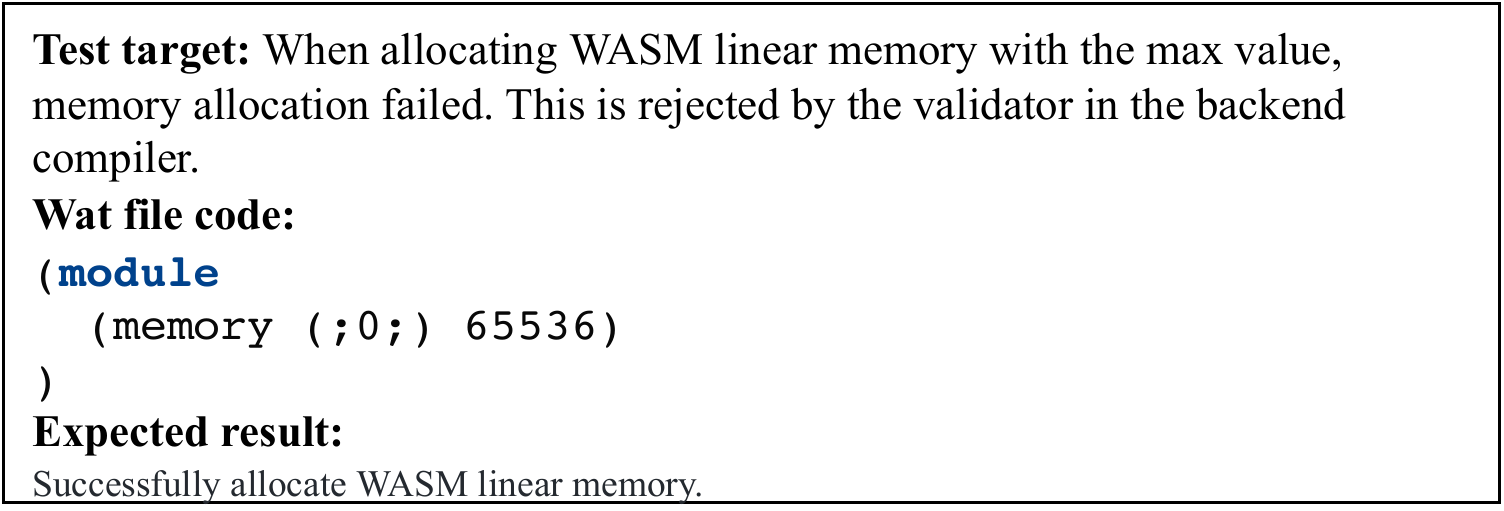}
	\caption{Case study of \texttt{[A.8] Validation error}}
	\label{figure:exampleF}
\end{figure}

As shown in Figure~\ref{figure:exampleF}, it is expected to allocate the linear memory to the max value. However, the allocation fails in \texttt{WAMR} and \texttt{wasm3}. This bug belongs to \textit{Validation error (A.8)} since the max value is not permitted by the validator. The developers in \texttt{WAMR} updated the max memory page value in the interpreter, and the developers from \texttt{wasm3} updated the max linear memory pages from 32768 to 65535 in the commit fbbacefeaf28e019244bbfa281fc4dea3dbdedc9.
Besides, it is expected to print the v128 data type to support the SIMD instructions. However, it prints nothing in \texttt{WasmEdge}. This bug belongs to \textit{Unsupported data operation (A.8)}. The developers have confirmed it is a bug and fixed it in the main branch.

\begin{figure}[htb]
	\includegraphics[width=0.75\textwidth]{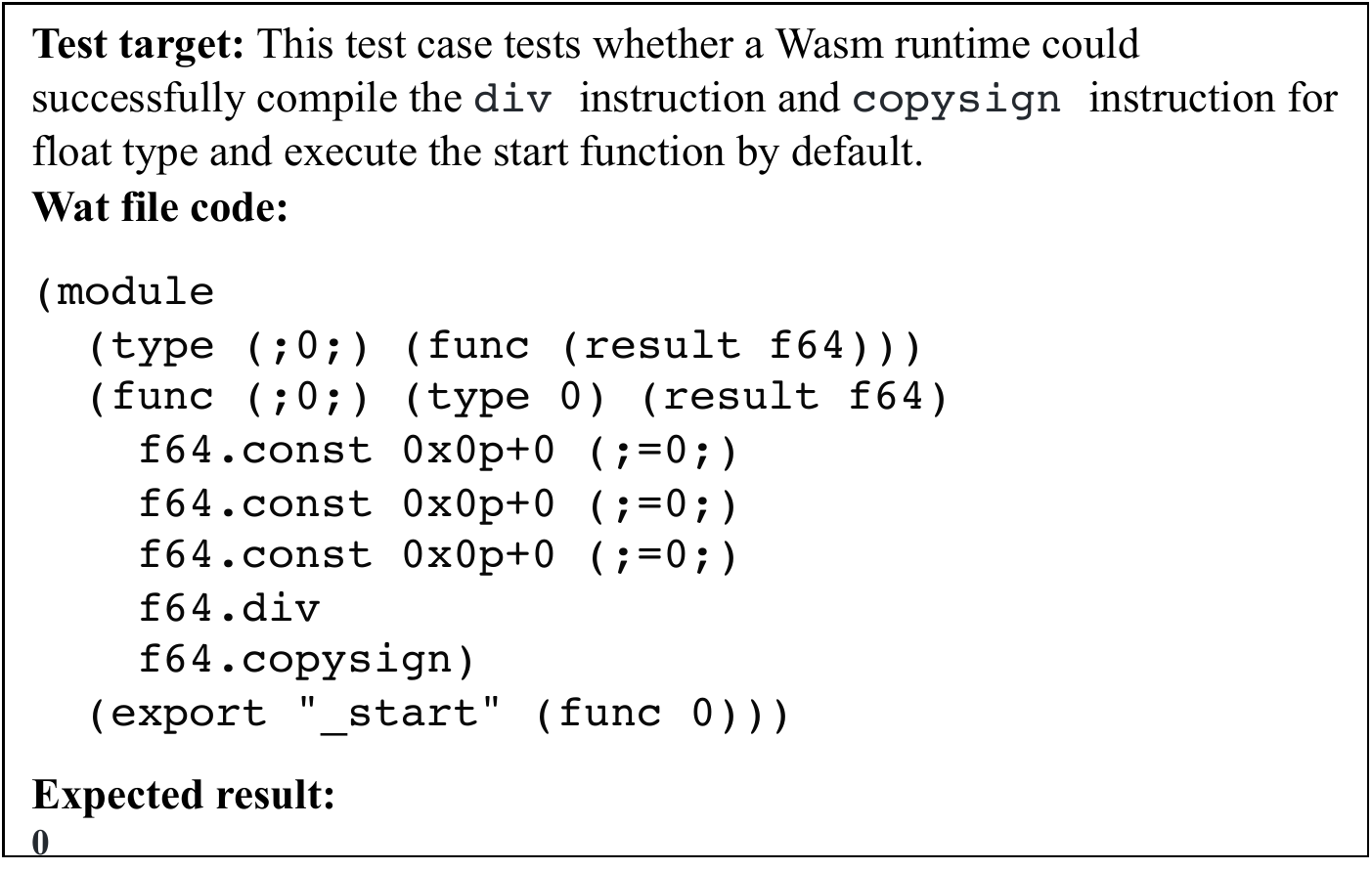}
	\caption{Case study of \texttt{[C.9] Entry point error}}
	\label{figure:example1}
\end{figure}

\begin{figure}[htb]
	\includegraphics[width=0.75\textwidth]{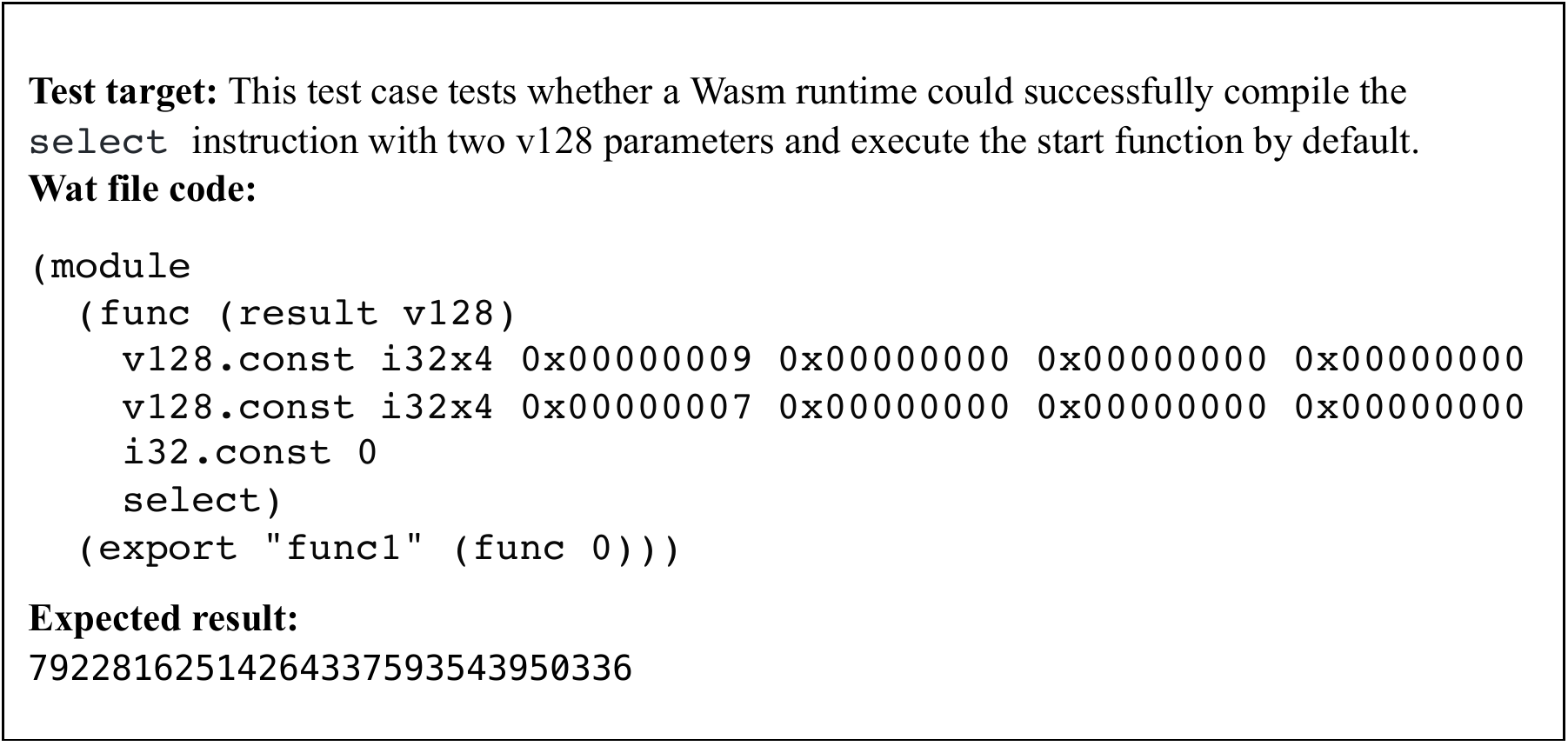}
	\caption{Case study of \texttt{[A.7] Unsupported data operation}}
	\label{figure:example3}
\end{figure}

Interestingly, we found that the test cases in our detection framework can trigger more than one types of bugs. 
For Example, the WASM module in Figure~\ref{figure:example1} is used to test whether WASM runtimes could successfully compile the \textit{div} and \textit{copysign} instructions for float data (\textit{[A.3] Compilation failure}). Beyond this, we found that it can identify bugs that belong to \textit{[C.9] Entry point error} in \texttt{wasm3}. The WASM module could be successfully compiled in \texttt{wasm3}. However, `\_start' is not considered the entry point in wasm3, although other WASM runtimes do. The developer confirmed it and considered fixing it by checking the return type of '\_start' and acting according to it. Moreover, the WASM module in Figure~\ref{figure:example3} is used to test whether WASM runtimes could successfully compile the \textit{select} instruction with two v128 parameters (\textit{[A.3] Compilation failure}). It detects a bug in \text{wasm3} which should be summarized to \textit{[A.7] Unsupported data operation}. Because WasmEdge could compile the module, it does not support printing v128 data. The developer confirmed that they only support print i32, i64, f32, and f64, which posed a bug, and they would fix it in the future.

\begin{tcolorbox}
\textbf{Highlight 8:} 
Our crafted bug detector can effectively detect bugs in real-world WASM runtimes and provide helpful information to facilitate bug diagnosis and fixing. Interestingly, we found that the test cases in our detection framework can trigger more than one types of bugs. It further suggests that the summarized bugs show similar patterns among different WASM runtimes.
\end{tcolorbox}

\section{Discussion}\label{sec:validity}
\subsection{Implications}

Given the rapidly increasing popularity of WASM, our study has timely and practical implications for both WASM runtime developers and researchers. 
First, our contribution could help developers dive into and resolve common bugs in WASM runtimes more efficiently. Our proposed bug detection framework could effectively detect bugs and provide useful information to facilitate bug fixing.
Second, as an emerging research direction, our study sheds lights on future studies on WASM, including automated testing of WASM runtimes, and developing more advance techniques to fix bugs, etc.

\subsection{Threats to Validity}
First, our analysis pipeline involves a manual analysis of bugs, which might introduce bias to our observations. 
To lower the influence of subjective threat, three authors take part in the analysis of bug and fix strategy analysis, discussing the inconsistent issues until reaching an agreement. 
Second, our empirical study only targets the most popular WASM runtimes, while there are many WASM runtimes in the wild, and they may pose other kinds of bugs that we did not cover in this paper. Nevertheless, we believe the selected three projects are representative enough for us to characterize common kinds of runtime bugs. 
Third, it is difficult to ensure that our crafted bug detectors are sound and can cover all the bug patterns of WASM runtimes. 
To deal with the problem, we perform a reliability evaluation of the bug detector and show that they can indeed trigger the known WASM runtime bugs. Nevertheless, for some bug reports, we cannot reproduce them to trigger the bugs the authors mentioned. We believe that some advanced techniques like fuzzing and differential testing can be adopted for complementary.

\section{Related Work}\label{sec:related}

\textbf{WebAssembly Runtime.}
WASM runtime has been used in a wide spectrum of applications. M{\'e}n{\'e}trey et al. proposed WebAssembly trusted runtime, TWINE \cite{twine}, to execute unmodified, language-independent applications. They leverage Intel SGX to build the runtime environment. Gadepalli et al.~\cite{sledge} proposed a light-weight WASM runtime, Sledge, for the edge computing. Wen et al. propose Wasmachine \cite{wasmachine2020}, an OS aiming to efficiently and securely execute WebAssembly applications in IoT and Fog devices with constrained resources. WASM runtime is the fundamental part of various applications. However, there are no studies about bugs in WASM runtimes. We present the first comprehensive study on characterizing and detecting bugs in WASM runtimes.

\textbf{Other WASM Related Studies.}
WASM is a promising and newly emerged area. There have been studies on several aspects of WASM, including the WASM execution efficiency \cite{herrera2018webassembly, jangda2019not, arewethere, bringing}, WASM compilers \cite{empiricalstudy_wasmcompiler, reverse, holk2018schism}, WASM binary security \cite{empiricalstudy_wasmbinaries, newagain, mcfadden2018security, reverse, wasm_code_obfuscation, lehmann2022finding}, etc. As WASM runtime is one of the fundamental components, our study provides timely insights to all stakeholders in the ecosystem.

\textbf{Empirical Study on bugs.}
There have been a large number of empirical studies focusing on software bugs across a wide range of applications. For example, Chen et al. \cite{empirical_DLdeployment} studied the faults related to the deployment of DL models on mobile devices; Zhang et al. \cite{empiricalstudy_TFbugs} conducted an empirical study of TensorFlow program bugs, Lu et al. \cite{lu2008learning} provided the comprehensive real-world concurrency bug characteristic study, Di Franco et al. \cite{di2017comprehensive} presented the first comprehensive study of real-world numerical bugs.
Recently, the rapid development of WebAssembly has inspired empirical studies on WebAsssmebly binaries and compilers. For example, Romano et al. \cite{empiricalstudy_wasmcompiler} conducted an empirical study of bugs in WebAssembly compilers. They investigated 146 bug reports in Emscripten related to the unique challenges WebAssembly compilers encounter compared with traditional compilers. Moreover, Hilbig et al. \cite{empiricalstudy_wasmbinaries} presented a comprehensive empirical study of 8,461 unique WebAssembly binaries. 
Following the widely used bug-studying method in the prior studies, we apply these bug characterization methods to the bugs in a different domain, i.e., WebAssembly runtime. Furthermore, we construct a bug detector for these summarized bugs based on the characterization and find 14 confirmed bugs.

\section{Conclusion}\label{sec:conclusion}
This paper has presented the first comprehensive study of bugs and the corresponding fix strategies of WASM runtimes. By manually analyzing 311 real-world bugs extracted from the most popular WASM runtimes, we have constructed a taxonomy of bug symptoms with 31 categories, and distilled the fix strategies for them. 
Based on the knowledge extracted, we further develop a pattern-based bug detection framework to automatically detect bugs across WASM runtimes. By the time of this study, we have identified 53 bugs that have never been reported in the community, and 14 of them have been confirmed by the official developers.

\bibliographystyle{ACM-Reference-Format}
\bibliography{sample-base}

\begin{thebibliography}{10}

\bibitem{reverse}
Reverse engineering webassembly.
\newblock \url{https://www.pnfsoftware.com/reversing-wasm.pdf}, 2018.

\bibitem{wasmer-go}
A complete and mature webassembly runtime for go based on wasmer.
\newblock \url{https://github.com/wasmerio/wasmer-go}, 2022.

\bibitem{craneliftdoc}
Cranelift doc.
\newblock
  \url{https://hacks.mozilla.org/2020/10/a-new-backend-for-cranelift-part-1-instruction-selection/},
  2022.

\bibitem{eos-vm}
Eos vm - a low-latency, high performance and extensible webassembly engine.
\newblock \url{https://github.com/EOSIO/eos}, 2022.

\bibitem{githubsearchapi}
Github search api.
\newblock \url{https://docs.github.com/cn/rest/search}, 2022.

\bibitem{hera}
hera - an ewasm (revision 4) virtual machine implemented in c++ conforming to
  evmc abiv9.
\newblock \url{https://github.com/ewasm/hera}, 2022.

\bibitem{life}
life - a secure and fast webassembly vm.
\newblock \url{https://github.com/perlin-network/life}, 2022.

\bibitem{Lucet}
Lucet - a native webassembly compiler and runtime.
\newblock \url{https://github.com/bytecodealliance/lucet}, 2022.

\bibitem{test_framework}
Test framework.
\newblock
  \url{https://drive.google.com/file/d/1XwgnL6F-oBwNBl-XOKc3h--JqFK-AwHQ/view?usp=sharing},
  2022.

\bibitem{wabt}
Wabt: The webassembly binary toolkit.
\newblock \url{https://github.com/WebAssembly/wabt}, 2022.

\bibitem{wagon}
wagon - a webassembly-based interpreter in go, for go.
\newblock \url{https://github.com/go-interpreter/wagon}, 2022.

\bibitem{wasilink}
Wasi link.
\newblock \url{https://wasi.dev/}, 2022.

\bibitem{wasm_non_web}
Wasm non web usage.
\newblock \url{https://webassembly.org/docs/non-web/}, 2022.

\bibitem{wasm_runtime_architecture_link}
Wasm runtime architecture.
\newblock \url{https://medium.com/wasm/webassembly-wasm-runtimes-522bcc7478fd},
  2022.

\bibitem{wasm3}
wasm3 - the fastest webassembly interpreter, and the most universal runtime.
\newblock \url{https://github.com/wasm3/wasm3}, 2022.

\bibitem{Wasm-Edge}
Wasmedge runtime.
\newblock \url{https://github.com/WasmEdge/WasmEdge}, 2022.

\bibitem{wasmer}
wasmer - a fast and secure webassembly runtime.
\newblock \url{https://github.com/wasmerio/wasmer}, 2022.

\bibitem{wasmerdoc}
Wasmer docs.
\newblock \url{https://docs.wasmer.io/}, 2022.

\bibitem{wasmi}
wasmi - webassembly (wasm) interpreter.
\newblock \url{https://github.com/paritytech/wasmi}, 2022.

\bibitem{wasmtime}
wasmtime - a standalone runtime for webassembly.
\newblock \url{https://github.com/bytecodealliance/wasmtime}, 2022.

\bibitem{wastfile}
Wast file.
\newblock
  \url{https://hacks.mozilla.org/2020/10/a-new-backend-for-cranelift-part-1-instruction-selection/},
  2022.

\bibitem{watfile}
Wat file.
\newblock
  \url{https://developer.mozilla.org/en-US/docs/WebAssembly/Text_format_to_wasm},
  2022.

\bibitem{WAVM}
Wavm - a webassembly virtual machine, designed for use in non-browser
  applications.
\newblock \url{https://github.com/WAVM/WAVM}, 2022.

\bibitem{WAMR}
Webassembly micro runtime.
\newblock \url{https://github.com/bytecodealliance/wasm-micro-runtime}, 2022.

\bibitem{wasidoc}
Webassembly system interface doc.
\newblock
  \url{https://hacks.mozilla.org/2019/03/standardizing-wasi-a-webassembly-system-interface/},
  2022.

\bibitem{wasmdoc}
Webassmebly doc.
\newblock \url{https://webassembly.org/}, 2022.

\bibitem{Emscripten}
Emscripten compiler.
\newblock \url{https://emscripten.org/}, 2023.

\bibitem{software_doc_issues}
{\sc Aghajani, E., Nagy, C., Vega-M{\'a}rquez, O.~L., Linares-V{\'a}squez, M.,
  Moreno, L., Bavota, G., and Lanza, M.}
\newblock Software documentation issues unveiled.
\newblock In {\em 2019 IEEE/ACM 41st International Conference on Software
  Engineering (ICSE)\/} (2019), IEEE, pp.~1199--1210.

\bibitem{wasm_blockchain}
{\sc Belchior, R., Vasconcelos, A., Guerreiro, S., and Correia, M.}
\newblock A survey on blockchain interoperability: Past, present, and future
  trends.
\newblock {\em ACM Computing Surveys (CSUR) 54}, 8 (2021), 1--41.

\bibitem{auto_classify_SO}
{\sc Beyer, S., Macho, C., Di~Penta, M., and Pinzger, M.}
\newblock Automatically classifying posts into question categories on stack
  overflow.
\newblock In {\em 2018 IEEE/ACM 26th International Conference on Program
  Comprehension (ICPC)\/} (2018), IEEE, pp.~211--21110.

\bibitem{wasm_code_obfuscation}
{\sc Bhansali, S., Aris, A., Acar, A., Oz, H., and Uluagac, A.~S.}
\newblock A first look at code obfuscation for webassembly.
\newblock In {\em Proceedings of the 15th ACM Conference on Security and
  Privacy in Wireless and Mobile Networks\/} (2022), pp.~140--145.

\bibitem{empirical_DLdeployment}
{\sc Chen, Z., Yao, H., Lou, Y., Cao, Y., Liu, Y., Wang, H., and Liu, X.}
\newblock An empirical study on deployment faults of deep learning based mobile
  applications.
\newblock In {\em 2021 IEEE/ACM 43rd International Conference on Software
  Engineering (ICSE)\/} (2021), IEEE, pp.~674--685.

\bibitem{kappa_journal}
{\sc Cohen, J.}
\newblock A coefficient of agreement for nominal scales.
\newblock {\em Educational and psychological measurement 20}, 1 (1960), 37--46.

\bibitem{real-world_numerical_bug}
{\sc Di~Franco, A., Guo, H., and Rubio-Gonz{\'a}lez, C.}
\newblock A comprehensive study of real-world numerical bug characteristics.
\newblock In {\em 2017 32nd IEEE/ACM International Conference on Automated
  Software Engineering (ASE)\/} (2017), IEEE, pp.~509--519.

\bibitem{di2017comprehensive}
{\sc Di~Franco, A., Guo, H., and Rubio-Gonz{\'a}lez, C.}
\newblock A comprehensive study of real-world numerical bug characteristics.
\newblock In {\em 2017 32nd IEEE/ACM International Conference on Automated
  Software Engineering (ASE)\/} (2017), IEEE, pp.~509--519.

\bibitem{ossbug}
{\sc Ding, Z.~Y., and Le~Goues, C.}
\newblock An empirical study of oss-fuzz bugs.
\newblock In {\em 2021 IEEE/ACM 18th International Conference on Mining
  Software Repositories (MSR)\/} (2021), IEEE, pp.~131--142.

\bibitem{sledge}
{\sc Gadepalli, P.~K., McBride, S., Peach, G., Cherkasova, L., and Parmer, G.}
\newblock Sledge: A serverless-first, light-weight wasm runtime for the edge.
\newblock In {\em Proceedings of the 21st International Middleware
  Conference\/} (2020), pp.~265--279.

\bibitem{wasm_serverless_edge}
{\sc Gadepalli, P.~K., Peach, G., Cherkasova, L., Aitken, R., and Parmer, G.}
\newblock Challenges and opportunities for efficient serverless computing at
  the edge.
\newblock In {\em 2019 38th Symposium on Reliable Distributed Systems (SRDS)\/}
  (2019), IEEE, pp.~261--2615.

\bibitem{bringing}
{\sc Haas, A., Rossberg, A., Schuff, D.~L., Titzer, B.~L., Holman, M., Gohman,
  D., Wagner, L., Zakai, A., and Bastien, J.}
\newblock Bringing the web up to speed with webassembly.
\newblock In {\em Proceedings of the 38th ACM SIGPLAN Conference on Programming
  Language Design and Implementation\/} (2017), pp.~185--200.

\bibitem{herrera2018webassembly}
{\sc Herrera, D., Chen, H., Lavoie, E., and Hendren, L.}
\newblock Webassembly and javascript challenge: Numerical program performance
  using modern browser technologies and devices.
\newblock {\em University of McGill, Montreal: QC, Technical report
  SABLE-TR-2018-2\/} (2018).

\bibitem{empiricalstudy_wasmbinaries}
{\sc Hilbig, A., Lehmann, D., and Pradel, M.}
\newblock An empirical study of real-world webassembly binaries: Security,
  languages, use cases.
\newblock In {\em Proceedings of the Web Conference 2021\/} (2021),
  pp.~2696--2708.

\bibitem{holk2018schism}
{\sc Holk, E.}
\newblock Schism: A self-hosting scheme to webassembly compiler.
\newblock {\em Proceedings of the Scheme and Functional\/} (2018).

\bibitem{jangda2019not}
{\sc Jangda, A., Powers, B., Berger, E.~D., and Guha, A.}
\newblock Not so fast: Analyzing the performance of $\{$WebAssembly$\}$ vs.
  native code.
\newblock In {\em 2019 USENIX Annual Technical Conference (USENIX ATC 19)\/}
  (2019), pp.~107--120.

\bibitem{newagain}
{\sc Lehmann, D., Kinder, J., and Pradel, M.}
\newblock Everything old is new again: Binary security of $\{$WebAssembly$\}$.
\newblock In {\em 29th USENIX Security Symposium (USENIX Security 20)\/}
  (2020), pp.~217--234.

\bibitem{lehmann2022finding}
{\sc Lehmann, D., and Pradel, M.}
\newblock Finding the dwarf: recovering precise types from webassembly
  binaries.
\newblock In {\em Proceedings of the 43rd ACM SIGPLAN International Conference
  on Programming Language Design and Implementation\/} (2022), pp.~410--425.

\bibitem{lu2008learning}
{\sc Lu, S., Park, S., Seo, E., and Zhou, Y.}
\newblock Learning from mistakes: a comprehensive study on real world
  concurrency bug characteristics.
\newblock In {\em Proceedings of the 13th international conference on
  Architectural support for programming languages and operating systems\/}
  (2008), pp.~329--339.

\bibitem{wasmiot}
{\sc M{\"a}kitalo, N., Mikkonen, T., Pautasso, C., Bankowski, V., Daubaris, P.,
  Mikkola, R., and Beletski, O.}
\newblock Webassembly modules as lightweight containers for liquid iot
  applications.
\newblock In {\em International Conference on Web Engineering\/} (2021),
  Springer, pp.~328--336.

\bibitem{mcfadden2018security}
{\sc McFadden, B., Lukasiewicz, T., Dileo, J., and Engler, J.}
\newblock Security chasms of wasm.
\newblock {\em NCC Group Whitepaper\/} (2018).

\bibitem{Wasm_edge_computing}
{\sc Mendki, P.}
\newblock Evaluating webassembly enabled serverless approach for edge
  computing.
\newblock In {\em 2020 IEEE Cloud Summit\/} (2020), IEEE, pp.~161--166.

\bibitem{twine}
{\sc M{\'e}n{\'e}trey, J., Pasin, M., Felber, P., and Schiavoni, V.}
\newblock Twine: An embedded trusted runtime for webassembly.
\newblock In {\em 2021 IEEE 37th International Conference on Data Engineering
  (ICDE)\/} (2021), IEEE, pp.~205--216.

\bibitem{platformbug}
{\sc Paltenghi, M., and Pradel, M.}
\newblock Bugs in quantum computing platforms: an empirical study.
\newblock {\em Proceedings of the ACM on Programming Languages 6}, OOPSLA1
  (2022), 1--27.

\bibitem{empiricalstudy_wasmcompiler}
{\sc Romano, A., Liu, X., Kwon, Y., and Wang, W.}
\newblock An empirical study of bugs in webassembly compilers.
\newblock In {\em 2021 36th IEEE/ACM International Conference on Automated
  Software Engineering (ASE)\/} (2021), IEEE, pp.~42--54.

\bibitem{wasim}
{\sc Romano, A., and Wang, W.}
\newblock Wasim: Understanding webassembly applications through classification.
\newblock In {\em 2020 35th IEEE/ACM International Conference on Automated
  Software Engineering (ASE)\/} (2020), IEEE, pp.~1321--1325.

\bibitem{qualitative_methods}
{\sc Seaman, C.~B.}
\newblock Qualitative methods in empirical studies of software engineering.
\newblock {\em IEEE Transactions on software engineering 25}, 4 (1999),
  557--572.

\bibitem{slicing_of_wasm_binaries}
{\sc Sti{\'e}venart, Q., Binkley, D.~W., and De~Roover, C.}
\newblock Static stack-preserving intra-procedural slicing of webassembly
  binaries.
\newblock In {\em 2022 IEEE/ACM 44th International Conference on Software
  Engineering (ICSE)\/} (2022), IEEE, pp.~2031--2042.

\bibitem{arewethere}
{\sc Wang, W.}
\newblock Empowering web applications with webassembly: Are we there yet?
\newblock In {\em 2021 36th IEEE/ACM International Conference on Automated
  Software Engineering (ASE)\/} (2021), IEEE, pp.~1301--1305.

\bibitem{pythonbug}
{\sc Wang, Z., Bu, D., Sun, A., Gou, S., Wang, Y., and Chen, L.}
\newblock An empirical study on bugs in python interpreters.
\newblock {\em IEEE Transactions on Reliability\/} (2022).

\bibitem{wasmachine2020}
{\sc Wen, E., and Weber, G.}
\newblock Wasmachine: Bring iot up to speed with a webassembly os.
\newblock In {\em 2020 IEEE International Conference on Pervasive Computing and
  Communications Workshops (PerCom Workshops)\/} (2020), IEEE, pp.~1--4.

\bibitem{wen2021empirical}
{\sc Wen, J., Chen, Z., Liu, Y., Lou, Y., Ma, Y., Huang, G., Jin, X., and Liu,
  X.}
\newblock An empirical study on challenges of application development in
  serverless computing.
\newblock In {\em Proceedings of the 29th ACM Joint Meeting on European
  Software Engineering Conference and Symposium on the Foundations of Software
  Engineering\/} (2021), pp.~416--428.

\bibitem{empiricalstudy_DLapplications}
{\sc Zhang, T., Gao, C., Ma, L., Lyu, M., and Kim, M.}
\newblock An empirical study of common challenges in developing deep learning
  applications.
\newblock In {\em 2019 IEEE 30th International Symposium on Software
  Reliability Engineering (ISSRE)\/} (2019), IEEE, pp.~104--115.

\bibitem{wasmbook}
{\sc Zhang, X.}
\newblock {\em WebAssembly Principles and Core Technologies}.
\newblock China Machine Press, 2020.

\bibitem{empiricalstudy_TFbugs}
{\sc Zhang, Y., Chen, Y., Cheung, S.-C., Xiong, Y., and Zhang, L.}
\newblock An empirical study on tensorflow program bugs.
\newblock In {\em Proceedings of the 27th ACM SIGSOFT International Symposium
  on Software Testing and Analysis\/} (2018), pp.~129--140.

\bibitem{gccbug}
{\sc Zhou, Z., Ren, Z., Gao, G., and Jiang, H.}
\newblock An empirical study of optimization bugs in gcc and llvm.
\newblock {\em Journal of Systems and Software 174\/} (2021), 110884.

\end{thebibliography}

\end{document}